\documentclass[
aps,
prd,
twocolumn,
superscriptaddress,
floatfix,
nobibnotes,
nofootinbib,
]
{revtex4-2}
\usepackage[english]{babel}
\usepackage{caption}
\usepackage{subcaption}
\usepackage[]{hyperref}
\usepackage[utf8]{inputenc}
\usepackage{amsmath,amssymb,amsthm,amstext}
\usepackage{mathtools}
\usepackage[nameinlink,capitalise]{cleveref}
\usepackage{siunitx}
\usepackage{braket}
\usepackage{verbatim}
\usepackage{multirow}
\usepackage{dsfont}
\usepackage{siunitx}
\usepackage{float}
\usepackage{graphicx}
\usepackage{changebar}
\usepackage{graphicx}
\usepackage{dcolumn}
\usepackage{bm}
\usepackage{orcidlink}
\usepackage{xcolor}
\usepackage{slashed}
\usepackage{esvect}
\usepackage{crossreftools}
\usepackage{hyperref}
\usepackage{enumitem}
\usepackage[normalem]{ulem}
\usepackage{makecell}

\newcommand{\Eq}[1]{{\cref{#1}}}
\renewcommand{\eqref}[1]{\cref{#1}}

\newcommand{\Darmstadt}{
    Technische Universität Darmstadt, 
    Fachbereich Physik, 
    Institut für Kernphysik, 
    Theoriezentrum, 
    Schlossgartenstr. 2, 
    D-64289 Darmstadt, 
    Germany 
}

\newcommand{\HFHFDarmstadt}{
    Helmholtz Forschungsakademie Hessen für FAIR (HFHF),
    GSI Helmholtzzentrum für Schwerionenforschung,
    Campus Darmstadt, D-64289 Darmstadt, Germany
}

\hypersetup{
	colorlinks=true,
	citecolor=blue,
	citebordercolor=red,
	linktoc=all,
	linkcolor=blue,
	urlcolor=blue
}

\AddToHook{cmd/appendix/before}{
    
    \crefalias{section}{appendix}
    \crefalias{subsection}{appendix}
}

\crefformat{equation}{#2Eq.~(#1)#3}
\crefformat{appendix}{#2App.~#1#3}
\crefformat{section}{#2Sec.~#1#3}
\crefformat{table}{#2Tab.~#1#3}
\crefformat{figure}{#2Fig.~#1#3}

\crefmultiformat{section}{#2Secs.~#1#3}{ and~#2#1#3}{, #2#1#3}{ and~#2#1#3}
\crefmultiformat{appendix}{#2Apps.~#1#3}{ and~#2#1#3}{, #2#1#3}{ and~#2#1#3}

\crefrangeformat{equation}{#3Eqs.~(#1)#4 to #5(#2)#6}

\Crefformat{figure}{#2Fig.~#1#3}

\graphicspath{{./figures/}}

\makeatother

\begin{document}

\title{Interplay between inhomogeneous chiral and crystalline color-superconducting phases in the two-flavor NJL model}

\author{Chengfu Mu}
\email[]{muchengfu@huznu.edu.cn}
\affiliation{Department of Physics, Huzhou University, Huzhou 313000, Zhejiang, China }
\author{Hosein Gholami \orcidlink{0009-0003-3194-926X}}
\email[]{mohammadhossein.gholami@tu-darmstadt.de}
\affiliation{\Darmstadt}
\author{Michael Buballa \orcidlink{0000-0003-3747-6865}}
\email[]{michael.buballa@tu-darmstadt.de}
\affiliation{\Darmstadt}
\affiliation{\HFHFDarmstadt}

\begin{abstract}
We study the interplay between the chiral density wave (CDW) and the single-plane-wave Larkin-Ovchinnikov-Fulde-Ferrell (LOFF) phase of color-superconducting matter in 
two-flavor quark matter at vanishing and non-vanishing temperature $T$, quark number chemical potential $\mu$ and isospin chemical potential $\delta\mu$.
The analysis is performed within
the two-flavor Nambu--Jona-Lasinio (NJL) model in the chiral limit, using a three-momentum cutoff scheme.  Treating the CDW wave vector
$\vec{q}$ and the LOFF pair momentum $\vec{q}\,'$ as independent variational
parameters, we minimize the mean-field effective potential with respect to
both amplitudes and both wave vectors, without constraining their relative
orientation, and map out the $T$-$\mu$ and $\mu$-$\delta\mu$ phase diagrams for
a range of diquark couplings $G_D$. Our central result is that $\vec{q}$ and
$\vec{q}\,'$ are never simultaneously nonzero: inhomogeneous chiral and diquark
condensates do not coexist across the entire parameter range.
\end{abstract}

\date{\today}
\maketitle
\section{Introduction}\label{sec:intro}

Quantum chromodynamics (QCD) is expected to exhibit a remarkably rich phase structure once the thermodynamic axes of temperature ($T$) and baryon chemical potential ($\mu_B$) are explored.  At low $T$ and small $\mu_B$ the ground state is characterized by confinement and the spontaneous breaking of chiral symmetry, forming the familiar hadronic world.  Raising the temperature restores chiral symmetry and liberates color charges, leading to the quark--gluon plasma \cite{Rischke:2003mt}
that is now routinely produced in heavy--ion collisions\cite{Achenbach:2023pba,STAR:2016use,PHENIX:2018for,SADHU2025100140}.  Pushing the system instead to large baryon densities opens a third regime where attractive channels in QCD favor 
diquark Cooper pairing; bulk matter is then expected to organize in one of several color--superconducting (CSC) states \cite{Alford:2007xm,Nardulli:2002ai,Schmitt:2025cqi}.

From an astrophysical perspective this high--density domain is particularly relevant: the cores of massive neutron stars may reach several times nuclear saturation density 
while staying at low temperatures \cite{Steiner:2010fz,Lattimer:2014wci,Lattimer:2015nhk}. 
At moderately high densities where strange quarks are suppressed by their mass, a possible pairing pattern is the two-flavor superconducting (2SC) phase, where the Cooper pairs consist of up and down quarks.
On the other hand, neutron-star matter is in $\beta$--equilibrium and electrically neutral, leading to a non-vanishing difference
$\delta\mu=(\mu_u-\mu_d)/2$
between the chemical-potentials of the up and down quarks. 
In such an environment the pairing could be prohibited by  
considerable Fermi–surface stress \cite{Alford:2001zr}.  A textbook measure of this stress is the Chandrasekhar--Clogston limit \cite{Clogston:1962zz,Chandrasekhar:1962}, 
$|\delta\mu| = \Delta/\sqrt{2}$, relating the chemical-potential difference 
to the pairing gap $\Delta$.
In a weak-coupling expansion,
BCS pairing with zero total momentum becomes energetically disfavored beyond this limit.
However, a possible way to relax the tension and thereby to extend the pairing regime is to allow
the diquark pair to carry finite momentum, giving rise to the Larkin--Ovchinnikov–Fulde--Ferrell (LOFF) phase \cite{Fulde:1964zz,Larkin:1964zn,Casalbuoni:2003wh,Anglani:2013gfu}, where the diquark condensates are non-uniform in space.

A similar mechanism has also been discussed in the context of the chiral condensate 
$\langle\bar{\psi}\psi\rangle$, the order parameter for chiral-symmetry breaking at low $T$ and $\mu$.
Just as the chemical-potential difference $\delta\mu$ imposes stress on the quark-quark pairs, the baryon chemical potential $\mu_B$ imposes stress on the quark-antiquark pairs in $\langle\bar{\psi}\psi\rangle$, giving rise to chiral-symmetry restoration at large $\mu_B$. This again can be relaxed to some extent by the formation of spatially modulated (``inhomogeneous'') chiral condensates \cite{Nakano:2004cd,Nickel:2009ke,Buballa:2014tba}.

Both LOFF phases and inhomogeneous chiral condensates break translational invariance and therefore lie outside the range of homogeneous mean--field analyzes.  Most existing studies focus on either the inhomogeneous chiral sector or the inhomogeneous pairing sector but only rarely on their interplay.  Yet the two order parameters compete for the same quark degrees of freedom.
It is therefore an interesting question whether they exclude each other or can coexist or even re-enforce each other, possibly 
depending on the density, temperature, flavor asymmetry and on the diquark coupling strength.

The present work tackles this missing piece by analyzing simultaneously the simplest representatives of both families---a single plane--wave ansatz for the chiral condensate -- a so--called chiral density wave (CDW) -- and a single plane--wave ansatz for the color--superconducting gap (two--flavor LOFF).  We employ the two--flavor Nambu--Jona--Lasinio (NJL) model in the chiral limit, whose pointlike interaction captures the relevant symmetries of QCD and has proved successful at describing both chiral symmetry breaking and color superconductivity
\cite{Klevansky:1992qe,BUBALLA2005205}.
 Working within a consistent three--momentum cutoff scheme we derive the mean--field effective potential, pay special attention to medium--induced divergences, and map out the $T$--$\mu$ and $\mu$--$\delta\mu$ phase diagrams 
 for different diquark couplings
 $G_D$. 
 Here $\mu = \mu_B/3$ is the quark chemical potential.

In contrast to previous treatments \cite{Casalbuoni:2003wh,Sadzikowski:2006jq,PhysRevD.103.034030,He:2006vr,Fukushima:2007bj},  we treat the two modulation vectors $\vec{q}$ 
(for CDW) and $\vec{q'}$ (for LOFF) as independent variational parameters and minimize the free energy with respect to 
both wave vectors (including their relative angle) and the corresponding amplitudes. 
This strategy allows us to establish clear criteria for when the CDW and LOFF phases meet, whether they can coexist, and how their competition reshapes the phase diagram relevant for compact stars.

The remainder of the paper is organized as follows. In \cref{sec:njl} we introduce the NJL Lagrangian with quark-antiquark and diquark channels, define the mean--field ansatz, and derive the mean-field effective potential in \cref{sec:eff}.
In \cref{regcutoff} we discuss the regularization and subtraction procedure used to tame ultraviolet divergences.  In \cref{Tmuphase} we present the $T$--$\mu$ phase diagram, while in \cref{mudeltamuphase} we analyze the $\mu$--$\delta\mu$ plane at zero temperature for several values of $G_D$.  We conclude in \cref{s5} with a summary and an outlook.

\section{The Model}
\label{sec:model}

\subsection{The NJL model with diquarks}
\label{sec:njl}
We consider the extended version of the NJL Lagrangian \cite{Fujihara:2008as,Huang:2001yw,Huang:2002zd,He:2006vr} 
\begin{equation}
\label{eq:Lagr}
 \mathcal{L}_{\textrm{NJL+}\Delta} = \mathcal{L}_\textrm{NJL} + \mathcal{L}_\Delta\,,
\end{equation}
where
\begin{equation}
\label{eq:Lagr_NJL}
    \mathcal{L}_{\textrm{NJL}} = \bar{\psi} (i \gamma^\mu \partial_\mu + \hat{\mu} \gamma_0 ) \psi + G_S \left[ (\bar{\psi} \psi)^2 + (\bar{\psi} i \gamma_5 \vec{\tau} \psi)^2 \right]
\end{equation}
is the standard Lagrangian of the NJL model for
$N_{f}=2$ quark flavors and $N_{c}=3$
color degrees of freedom in the chiral limit and at
finite chemical potential $\hat{\mu}$, where $\hat{\mu} = \text{diag}(\mu_u,\mu_d)$.
$\psi$ is a $4N_{c}N_{f}$-dimensional quark spinor, $G_S$ is the scalar four-fermion coupling, $\gamma^{\mu}$ are the Dirac matrices, and $\vec{\tau} = (\tau_1, \tau_2, \tau_3)$ represent the Pauli matrices in flavor space.

The second term in \cref{eq:Lagr} is introduced to describe diquark condensation in the spin-zero color-antitriplet channel
\begin{equation}
    \mathcal{L}_{\Delta} = G_{D} \left(\bar{\psi}_c i \gamma_5 \varepsilon  \epsilon^{b} \psi\right) \left(\bar{\psi} i \gamma_5 \varepsilon  \epsilon^{b} \psi_c\right) \,,
\end{equation}
where $\psi_c = C \bar{\psi}^{\textrm{T}}$ and $\bar{\psi}_c = \psi^{\textrm{T}} C$, with $C = i \gamma^2 \gamma^0$ being the charge-conjugation matrix, and $(\varepsilon)^{ik} \equiv \varepsilon^{ik}$,
$(\epsilon^b)^{\alpha \beta} \equiv \epsilon^{\alpha \beta b}$ are totally
antisymmetric tensors in the flavor and color spaces, respectively.

The order parameter for the 2SC phase is defined as
\begin{align}
\Delta^b =&
2G_{D}\left\langle\bar{\psi}_c i\gamma_5\varepsilon  \epsilon^{b}\psi\right\rangle, 
\end{align}
with color index $b$. 
In the following we assume that only the $b=3$ component (``anti-blue'') is non-vanishing,
\begin{align}
\label{meanfield}
  \Delta^1 =   \Delta^2  = 0,\,  \Delta^3 \neq 0,
\end{align}
which can always be achieved by a (local) color rotation. This means that only the 
first two colors (``red'' and ``green'' by convention)
participate in the condensate, while the 
third (``blue'')
does not. In the
later expressions, we will simply write $\Delta \equiv \Delta^3$.

For the chiral phase, the order parameters are given by the scalar and pseudoscalar condensates
\begin{align}
\sigma =& -2G_{S}\left\langle\bar{\psi}\psi\right\rangle, \\
\pi_a =& -2G_{S}\left\langle\bar{\psi}i\gamma_5\tau_a\psi\right\rangle \quad (a = 1, 2, 3).
\end{align}

Under the mean-field approximation, the effective Lagrangian then takes the following form
\begin{align}
\label{eq:Leff1}
    \mathcal{L}_{\textrm{eff}} = \dfrac{1}{2}  & \Bigg[ \bar{\psi} (i  \slashed{\partial} + \hat{\mu} \gamma_0 - \sigma(x)
   -  i \gamma_5 \vec{\pi}(x) \! \cdot \! \vec{\tau}) \psi \nonumber\\
    &+ \bar{\psi}_c (i \slashed{\partial} - \hat{\mu} \gamma_0 -  \sigma(x)
   -  i \gamma_5 \vec{\pi}(x) \! \cdot \! \vec{\tau}^T) \psi_c \notag \\
    &+ \Delta^*(x) \left(\bar{\psi}_c i \gamma_5 \varepsilon  \epsilon^{3}\psi\right) + \Delta(x) \left(\bar{\psi} i \gamma_5 \varepsilon  \epsilon^{3} \psi_c\right) \nonumber\\
    &- \dfrac{\sigma^2(x) + \vec{\pi}^2(x)}{2 G_S} - \dfrac{|\Delta(x)|^2}{2 G_D} \Bigg].    
\end{align}
Here, as indicated, both the chiral and the color superconducting mean fields are in general space-time dependent quantities.
In this form, the evaluation of the effective potential is extremely difficult, even in mean-field approximation. Therefore further simplifications are needed in order to progress.
In the following we assume the condensates to be time-independent and adopt a single-plane-wave ansatz for their spatial dependence.  
More precisely, in the chiral sector, we make the CDW ansatz\footnote{Note that some authors call this ansatz a ``dual chiral density wave'' (DCDW) because both the scalar and the pseudoscalar condensate are modulated, see, e.g., Ref.~\cite{Nakano:2004cd}.}
\begin{equation}
\sigma(x) = M \cos{\vec{q}\cdot\vec{x}}, \quad 
\pi_a(x) =  M \delta_{a3} \sin{\vec{q}\cdot\vec{x}},
\end{equation}
with wave vector $\vec{q}$.
The combination of the two fields
can be expressed as
\begin{equation}
\sigma(x) + i\gamma_5 \tau_3 \pi_3(x) =  M \exp({i\gamma_5 \tau_3\vec{q}\cdot\vec{x}}),
\end{equation}
which will turn out to be advantageous.
Also note that the CDW ansatz contains the standard homogeneous chiral condensate $\sigma = M = \mathit{const.}$ as the special case $M \neq 0$ and $\vec{q} = 0$.

Similarly, we adopt the single-plane-wave ansatz for the diquark condensate
\begin{equation}
\Delta(x) = \Delta e^{2i\vec{q^\prime}\cdot\vec{x}}, \quad \Delta^*(x) = \Delta e^{-2i\vec{q^\prime}\cdot\vec{x}},
\end{equation}
where $\vec{q^\prime}$ represents the LOFF pair momentum.\footnote{This ansatz is sometimes referred to as Fulde-Ferrell (FF) phase.}
Again, the homogeneous 2SC phase is contained in this ansatz as the special case $\Delta \neq 0$ and $\vec{q'} = 0$.

A general difficulty with inhomogeneous phases is that  the quark propagator contains off-diagonal terms in momentum space. However, the single-plane-wave ans{\"a}tze above have the advantage that they allow for an analytic diagonalization of the momentum-space part. The most elegant way to do this is by employing 
two consecutive field transformations. 

The first step involves a chiral field transformation
along with its conjugates~\cite{Dautry:1979bk,KUTSCHERA1990566}
\begin{alignat}{2}
\label{eq:transf1}
&\psi(x) &&= \exp\left({-\dfrac{i}{2}\gamma_5 \tau_3\vec{q}\cdot\vec{x}}\right)\psi^\prime(x), \notag \\
&\psi_c(x) &&= \exp\left({-\dfrac{i}{2}\gamma_5 \tau_3\vec{q}\cdot\vec{x}}\right)\psi^\prime_c(x), \notag \\
&\bar{\psi}(x) &&= \bar{\psi}^\prime(x)\exp\left({-\dfrac{i}{2}\gamma_5 \tau_3\vec{q}\cdot\vec{x}}\right), \notag \\
&\bar{\psi}_c(x)&&= \bar{\psi}^\prime_c(x)\exp\left({-\dfrac{i}{2}\gamma_5 \tau_3\vec{q}\cdot\vec{x}}\right).
\end{alignat}
Inserting this into \Eq{eq:Leff1}
yields the following effective Lagrangian
\begin{align}
    \mathcal{L}_{\textrm{eff}} = \dfrac{1}{2}  & \Bigg[ \bar{\psi}^\prime (i  \gamma^\mu \partial_\mu-\dfrac{1}{2}\gamma_5 \tau_3 \vec\gamma\cdot\vec{q} + \hat{\mu} \gamma_0 -M) \psi^\prime \nonumber\\
    &+ \bar{\psi}^\prime_c (i \gamma^\mu \partial_\mu-\dfrac{1}{2}\gamma_5 \tau_3 \vec\gamma\cdot\vec{q} - \hat{\mu} \gamma_0-M) \psi^\prime_c \notag \\
    &+ \Delta^*(x) \left(\bar{\psi}^\prime_c i \gamma_5 \varepsilon  \epsilon^{3} \psi^\prime\right) 
    + \Delta(x) \left(\bar{\psi}^\prime i \gamma_5 \varepsilon  \epsilon^{3} \psi^\prime_c\right) \nonumber\\
    &- \dfrac{M^2}{2 G_S} - \dfrac{|\Delta|^2}{2 G_D} \Bigg] ,
\end{align}
where the space dependent fields $\sigma$ and $\pi_3$ have been transformed away to the expense of the easier-to-handle terms proportional to $\vec{q}$. Note that the quark-quark-diquark terms (third row) are invariant under the chiral transformation because $\varepsilon \equiv i\tau_2$ anticommutes with $\tau_3$. 

Next we also remove the space dependent diquark fields by a second transformation~
\cite{He:2006vr,Bowers:2001ip},

\begin{alignat}{2}
\label{eq:transf2}
&\psi^\prime(x) &&= \chi(x)e^{i\vec{q^\prime}\cdot\vec{x}}, \notag \\
&\psi^\prime_c(x) &&= \chi_c(x)e^{-i\vec{q^\prime}\cdot\vec{x}}, \notag \\
&\bar{\psi}^\prime(x) &&= \bar{\chi}(x)e^{-i\vec{q^\prime}\cdot\vec{x}}, \notag \\
&\bar{\psi}^\prime_c(x) &&= \bar{\chi}_c(x)e^{i\vec{q^\prime}\cdot\vec{x}}.
\end{alignat}
In terms of these fields,
the effective Lagrangian reads
\begin{align}
    \mathcal{L}_{\textrm{eff}}
       =& \dfrac{1}{2}  \Bigg[ \bar{\chi}(x) \left(i  \gamma^\mu \partial_\mu-\dfrac{1}{2}\gamma_5 \tau_3 \vec{\gamma}\cdot\vec{q} -\vec{\gamma}\cdot\vec{q^\prime} + \hat{\mu} \gamma_0 -M \right) \chi(x) \notag \\
    &+ \bar{\chi}_c(x) \left(i \gamma^\mu \partial_\mu-\dfrac{1}{2}\gamma_5 \tau_3 \vec{\gamma}\cdot\vec{q}  +\vec{\gamma}\cdot\vec{q^\prime} - \hat{\mu} \gamma_0 -M \right) \chi_c(x) \notag \\
    &+ \Delta^* \left(\bar{\chi}_c(x) i \gamma_5 \varepsilon  \epsilon^{3} \chi(x)\right) 
    + \Delta \left(\bar{\chi}(x) i \gamma_5 \varepsilon  \epsilon^{3} \chi_c(x)\right) \notag \\
    &- \dfrac{M^2 }{2 G_S} - \dfrac{|\Delta|^2}{2 G_D} \! \Bigg]. 
\end{align}
Finally, we introduce the Nambu-Gor'kov spinor
\begin{equation}
    \Psi(x) = \left(
    \begin{array}{c}
         \chi(x)  \\
         \chi_c(x) 
    \end{array}
    \right),
\end{equation}
allowing us to write the effective Lagrangian in the compact form 
\begin{equation}
\label{effective Lagrangian}
    \mathcal{L}_{\textrm{eff}}
    =\frac{1}{2} \bar{\Psi}\mathcal{S}^{-1}\Psi- \dfrac{M^2 }{4 G_S} - \dfrac{|\Delta|^2}{4 G_D},    
\end{equation}
where
\begin{widetext}
\begin{align}
    \mathcal{S}^{-1}& = \begin{pmatrix} i \gamma^\mu \partial_\mu-\dfrac{1}{2}\gamma_5 \tau_3 \vec\gamma\cdot\vec{q} -\vec{\gamma}\cdot\vec{q^\prime} - M + \hat{\mu} \gamma^0 & \Delta i \gamma_5 \varepsilon  \epsilon^{3} \\
     \Delta^* i \gamma_5 \varepsilon  \epsilon^{3}& i \gamma^\mu \partial_\mu-\dfrac{1}{2}\gamma_5 \tau_3 \vec\gamma\cdot\vec{q} +\vec{\gamma}\cdot\vec{q^\prime}-M - \hat{\mu} \gamma^0  \end{pmatrix}
\end{align}
\end{widetext}
is the inverse propagator in Nambu-Gor'kov space.

\subsection{Mean-field effective potential}
\label{sec:eff}

The effective potential per volume $V$, from which we later want to determine the phase structure of the model at temperature $T$ and chemical potentials $\mu_u = \mu + \delta\mu$ and  $\mu_d = \mu - \delta\mu$, 
can be derived within the path integral formalism as 
\begin{align}
\label{parthuang2-1}
\Omega^\text{eff}  = & -\frac{ T}{ V }{\rm ln}{\cal Z},
\end{align}
where
\begin{align}
\label{part}
{\cal Z} = N' \int [d {\bar{\Psi}} ][d \Psi] \exp\left\{ \int_0^{\beta} d \tau \int d^3{\vec x}
  ~ \mathcal{L}_{\textrm{eff}}\right\}
\end{align}
is the grand canonical partition function,
$\tau = it$ is the imaginary time variable, $\beta = 1/T$, and $N'$ is a normalization constant.  

Since the mean-field effective Lagrangian \eqref{effective Lagrangian} is bilinear in the (Nambu-Gor'kov) fermion fields, 
the path integral can be performed exactly. This yields
\begin{align}
\Omega^\text{eff}  = & -\frac{T}{V} \frac{1}{2}\mathrm{Tr}\ln(\beta \mathcal{S}^{-1}) + \dfrac{M^2 }{4 G_S} + \dfrac{|\Delta|^2}{4 G_D},
\end{align}
where $\mathrm{Tr}$ denotes a functional trace running over $V_4 =VT$ and internal (Dirac, color, flavor and Nambu-Gor'kov) degrees of freedom. 
The artificial doubling of the degrees of freedom in Nambu-Gor'kov formalism is compensated by the
factor of $\frac{1}{2}$.

After the field transformations \Eq{eq:transf1} and \Eq{eq:transf2} the inverse propagator is diagonal in momentum space. After a Fourier transformation the space-time part of the functional trace then simply becomes an integral over the quark three-momentum $\vec p$ and a sum over fermionic Matsubara frequencies $\omega_n = (2n +1)\pi T$, and we get
\begin{align}
\Omega^\text{eff}  = & -\int \frac{d^3p}{(2\pi)^3} \,T\sum\limits_n
 \frac{1}{2}\mathrm{tr}\ln(\beta \mathcal{S}^{-1}) + \dfrac{M^2 }{4 G_S} + \dfrac{|\Delta|^2}{4 G_D},
\end{align}
where the trace is now over internal degrees of freedom only.

Since the condensates are time-independent, the Matsubara sum can also be performed analytically. 
To this end we write
\begin{equation}
 \mathcal{S}^{-1}(i\omega_n,\vec p) =
 \gamma^0 (i\omega_n - \mathcal{H}(\vec p)),
\end{equation}
with the effective Dirac Hamiltonian $\mathcal{H}$, which does not depend on the Matsubara frequencies. 
Then, using 
\begin{equation}
  T \sum_{n}  \mathrm{ln} \left( \dfrac{i \omega_n + \mathcal{E}_i}{T} \right)  
  = \dfrac{\mathcal{E}_i}{2} + T \, \mathrm{ln} \left( 1 + \mathrm{e}^{-\mathcal{E}_i/T} \right).
\end{equation}
and the matrix identity $\mathrm{tr}\ln A = \ln\det A$, the effective potential can be brought into the following form
\begin{align}
\label{eq:omega_ev}
&\Omega^\text{eff}(T,\mu,\delta\mu,M,\Delta,\vec q,\vec q^\prime)\nonumber\\ &= \frac{M^2}{4G_S} + \frac{\Delta^2}{4G_D} \notag 
- \int \frac{d^3 p}{(2\pi)^3} \sum_{i=1}^{24}
\left[ \dfrac{\mathcal{E}_i}{2} + T \, \mathrm{ln} \left( 1 + \mathrm{e}^{-\mathcal{E}_i/T} \right) \right], 
\\
\end{align}
where $\mathcal{E}_i = \mathcal{E}_i(\vec p)$ are the eigenvalues of $\mathcal{H}$, corresponding to the dispersion relations of the quark quasiparticles in the presence of the condensates.

 We numerically deal with the eigenvalue spectrum of $\mathcal{H}$ in the following section. Here we just comment on the general structure. Taking into account the Dirac, color, flavor and Nambu-Gor'kov structure, the inverse propagator $ \mathcal{S}^{-1}$ and, hence, the Dirac Hamiltonian $\mathcal{H}$ are $48 \times 48$ matrices. There is, however, a two-fold degeneracy of each eigenvalue, so that the sum in \Eq{eq:omega_ev}
runs over 24 eigenvalues, corresponding to the 24 physical, i.e., four Dirac, three color and two flavor degrees of freedom.  

Eight of of them correspond to the blue quarks, which, according to our choice in \Eq{meanfield}, do not participate in the diquark pairing. We therefore simply recover the known dispersion relations of a non-superconducting CDW \cite{Dautry:1979bk} in this sector, 
\begin{align}
\label{dispersion1-2a}
\mathcal{E}_i
= 
E_\pm(\vec{p},M,\vec{q}) \pm (\mu \pm \delta\mu),
\end{align}
with 
\begin{align}
\label{dispersion1-2b}
E_\pm(\vec{p},M,\vec{q})
=
\sqrt{\vec{p}\,^2+M^2+\frac{1}{4}\vec{q}\,^2 \pm \sqrt{\vec{q}\,^2M^2+(\vec{p}\cdot \vec{q})^2}}.
\end{align}
The $\pm$ signs in \Eq{dispersion1-2a} can all be chosen independently, leading to eight dispersion relations, which we will identify with the eigenvalues $\mathcal{E}_{17}, \dots, \mathcal{E}_{24}$ in the following.
The signs in \Eq{dispersion1-2b} account for the splitting of energy levels due to the interaction of quark momenta $\vec{p}$ with the wave vector $\vec{q}$ of the CDW. For $\vec{q} = 0$, the expressions reduce to the simpler homogeneous case.

The remaining 16 eigenvalues correspond to the red and green quarks, which are paired in the 2SC condensate. In the limit of homogeneous pairing ($\vec{q'} = 0$) we reproduce the dispersion relations first derived in Ref.~\cite{Sadzikowski:2002iy}.
In the other limit of only LOFF without CDW, i.e., $\vec{q'} \neq 0$, $\vec{q} = 0$  and $M=0$, we can recover the known result obtained by He et al. \cite{He:2006vr}.
In general, however, i.e., for $\vec{q} \neq 0$ and $\vec{q'} \neq 0$, the eigenvalues in the red-green sector must be computed numerically by diagonalizing two $8\times 8$ matrices.

\subsection{Three-momentum hard cutoff regularization
\label{regcutoff}}

The integral in \Eq{eq:omega_ev} is divergent and needs to be regularized.
Various regularization schemes can handle this divergence, with each method impacting the resulting phase structure. For discussions of different regularization schemes in the context of inhomogeneous chiral phases, see Refs.~\cite{Broniowski:1990dy,Partyka:2008sv,Pannullo:2024sov}. A detailed comparison of regularization schemes, including renormalization group (RG)-consistent treatments, will be addressed in a future work.

In this work, we adopt the three-momentum cutoff scheme outlined in Refs.~\cite{Sadzikowski:2002iy,Sadzikowski:2006jq} for simplicity.
To this end, we first rearrange the temperature independent part of the eigenvalue sum in \Eq{eq:omega_ev} by adding and subtracting the blue-quark contributions ($i = 17, \dots, 24$),
\begin{align}
    - \sum_{i=1}^{24} \dfrac{\mathcal{E}_i}{2} 
    &=\, -3\sum_{i=17}^{24} \dfrac{\mathcal{E}_i}{2} 
    - \Big(\sum_{i=1}^{16} \dfrac{\mathcal{E}_i}{2} - 2\sum_{i=17}^{24} \dfrac{\mathcal{E}_i}{2}\Big)  
    \notag\\
    &= -6(E_+ + E_-) - \Big(\sum_{i=1}^{16} \dfrac{\mathcal{E}_i}{2} - 4(E_+ + E_-)\Big) , 
\end{align}
where in the second line we have readily performed the sum over the blue-quark eigenvalues, \Eq{dispersion1-2a}.
With this, the effective potential \Eq{eq:omega_ev} takes the form,
\begin{align}
\Omega^\text{eff}&(T,\mu,\delta\mu,M,\Delta,\vec q,\vec q^\prime)\nonumber\\ =& \frac{M^2}{4G_S} + \frac{\Delta^2}{4G_D} 
-6\int^\Lambda \frac{d^3 p}{(2\pi)^3} (E_+ + E_-)
\notag \\
& -\int^\Lambda \frac{d^3 p}{(2\pi)^3}\, \Big(\sum_{i=1}^{16} \dfrac{\mathcal{E}_i}{2} - 4(E_+ + E_-)\Big) 
\notag \\
&-\int^\Lambda \frac{d^3 p}{(2\pi)^3} 
\sum_{i=1}^{24}
 T \, \mathrm{ln} \left( 1 + \mathrm{e}^{-\mathcal{E}_i/T} \right) , 
\end{align}
where we indicated that we are going to regularize all integrals with a sharp three-momentum cutoff $\Lambda$. This includes the temperature dependent part (last line), even though it would be finite without regularization.

The integral in the second line is a pure vacuum contibution, i.e., it does not depend on $T$ or the chemical potentials. It  also does not depend on $\Delta$ or $\vec{q'}$. It depends, however, on the wave vector $\vec q$, giving rise to a term proportional to $\vec{q}\,^2 \Lambda^2$. 
Following Refs.~\cite{Sadzikowski:2002iy,Sadzikowski:2006jq} we expand this term in powers of $\vec q\,^2$,
\begin{align}
   -6\int^\Lambda &\frac{d^3 p}{(2\pi)^3}  (E_+ + E_-) 
   \notag\\
   &=
   -12\int^\Lambda \frac{d^3 p}{(2\pi)^3} E_p + \frac{M^2F_{\pi}^2}{2M_0^2}\vec{q}\,^2 + \dots,
\end{align}
and neglect all higher orders.
Here $E_p =\sqrt{\vec p\,^2 + M^2}$ is the on-shell energy of an unpaired quark in homogeneous matter, while $F_\pi$ and $M_0$ are the pion decay constant and the constituent quark mass in vacuum, respectively. Note that the coefficient of the $\vec q\,^2$ term is a model independent result~\cite{Sadzikowski:2002iy,Sadzikowski:2006jq,Broniowski:1990dy}. We thus have
\begin{align}
\Omega^\text{eff}&(T,\mu,\delta\mu,M,\Delta,\vec q,\vec q^\prime)\nonumber\\ =& \frac{M^2}{4G_S} + \frac{\Delta^2}{4G_D} + \frac{M^2F_{\pi}^2}{2M_0^2}\vec{q}\,^2
\notag \\
& -\int^\Lambda \frac{d^3 p}{(2\pi)^3}\, \Big(
12 E_p - 4(E_+ + E_-) + \sum_{i=1}^{16} \dfrac{\mathcal{E}_i}{2} \Big) 
\notag \\
&-\int^\Lambda \frac{d^3 p}{(2\pi)^3} 
\sum_{i=1}^{24}
 T \, \mathrm{ln} \left( 1 + \mathrm{e}^{-\mathcal{E}_i/T} \right) , 
\end{align}
which for a homogeneous 2SC phase ($\vec{q'}= 0$) reduces to the result of Ref.~\cite{Sadzikowski:2006jq}.

However, the LOFF ansatz still leads to the presence of unphysical terms proportional to $\Lambda^2 \vec{q^\prime}\,^2$~\cite{Fukushima:2007bj}. These terms result from the spatial asymmetry of the quasiparticle spectrum introduced by the three-momentum cutoff~\cite{PhysRevD.82.056006,Gorbar:2005ru}.
In this work, we therefore adopt the subtraction scheme from Ref.~\cite{He:2006vr} to eliminate these contributions. The effective potential for numerical calculations is then expressed as
\begin{align}
&\Omega_{\text{sub}}(T,\mu,\delta\mu,M,\Delta,\vec{q},\vec{q^\prime}) \nonumber\\
&\quad= \Omega^\text{eff}(T,\mu,\delta\mu,M,\Delta,\vec{q},\vec{q^\prime}) \nonumber\\
&\quad - \Omega^\text{eff}(T,\mu,\delta\mu,M=0,\Delta=0,\vec{q}=0,\vec{q^\prime}) \nonumber\\
&\quad + \Omega^\text{eff}(T,\mu,\delta\mu,M=0,\Delta=0,\vec{q}=0,\vec{q^\prime}=0).
\end{align}
The subtraction scheme can be explained by first removing all contributions with $\vec{q^\prime}$ nonzero and other arguments zero. The final addition restores the part corresponding to $\vec{q^\prime} = 0$. 

To determine the physical state, the effective potential $\Omega_{\text{sub}}$ is minimized with respect to the variables $M$, $\Delta$, $\vec{q}$, and $\vec{q^\prime}$ at finite temperature $T$, average chemical potential $\mu$ and chemical potential mismatch $\delta\mu$. The corresponding gap equations are
\begin{align}
\frac{\partial \Omega_{\text{sub}}}{\partial M} =& 0,  \\
\frac{\partial \Omega_{\text{sub}}}{\partial \Delta} =& 0,  \\
\frac{\partial \Omega_{\text{sub}}}{\partial q_i} =& 0 \quad (i = x, y, z), \label{gapeqq} \\
\frac{\partial \Omega_{\text{sub}}}{\partial q^\prime_j} =& 0 \quad (j = x, y, z)\label{gapeqqp} . 
\end{align}
Since, without loss of generality, we are free to choose our coordinate frame in such a way that $\vec q$ points to the $z$ direction and $\vec{q'}$ lies in the $xz$ plane, these are effectively five independent equations. The ground state for given $T$, $\mu$ and $\delta\mu$ is then given by the solution with minimum free energy.

The simultaneous solution of \cref{gapeqq,gapeqqp} ensures that no specific orientation is imposed on the wave vectors $\vec{q}$ and $\vec{q}\,'$, which is especially important in a coexistence phase with \(q \neq 0\) and \(q' \neq 0\). However, as we will see in the next chapter, we do not find a region in which CDW and LOFF modulations coexist.

\begin{table*}[t]
\begin{center}
\caption{Phases and corresponding order parameters studied in this paper.
 \label{table01}}
 \bgroup
\def\arraystretch{1.5}
\begin{tabular}{c>{\hspace{1pc}}c>{\hspace{1pc}}c>{\hspace{1pc}}c>{\hspace{1pc}}c>{\hspace{1pc}}l}
\hline\hline
phase   &  M  &  q  &  $\Delta$  &  $q^\prime$ & description\\
\hline
 R & 0     &0       &0           & 0& chiral-symmetry restored, non-superconducting \\
 Ch & $ \ne 0$     & 0       & 0           &0& chiral symmetry homogeneously broken\\
NCh   & $ \ne 0$     &$ \ne 0$      & 0           &0& chiral symmetry non-uniformly broken (CDW)\\
 2SC  & 0& 0 & $ \ne 0$  &0& homogeneous color superconductor \\
 LOFF & 0 & 0 & $ \ne 0$           &$ \ne 0$& non-uniform color superconductor  \\
 HC  & $ \ne 0$     & 0       & $ \ne 0$           &0& coexistence of homogeneous chiral and 2SC condensates \\
 SNCh & $ \ne 0$     & $ \ne 0$       & $ \ne 0$           &0& coexistence of a CDW with homogeneous 2SC\\
(not observed) & $ \ne 0$     & $0$       & $ \ne 0$           & $\ne 0$& coexistence of LOFF with a homogeneous chiral condensate\\
(not observed) & $ \ne 0$     & $ \ne 0$       & $ \ne 0$           & $\ne 0$& coexistence of LOFF with a CDW\\
\hline\hline
\end{tabular}
\egroup
\end{center}
\end{table*}

\begin{figure*}[t]
\begin{center}
\includegraphics[width=8cm]{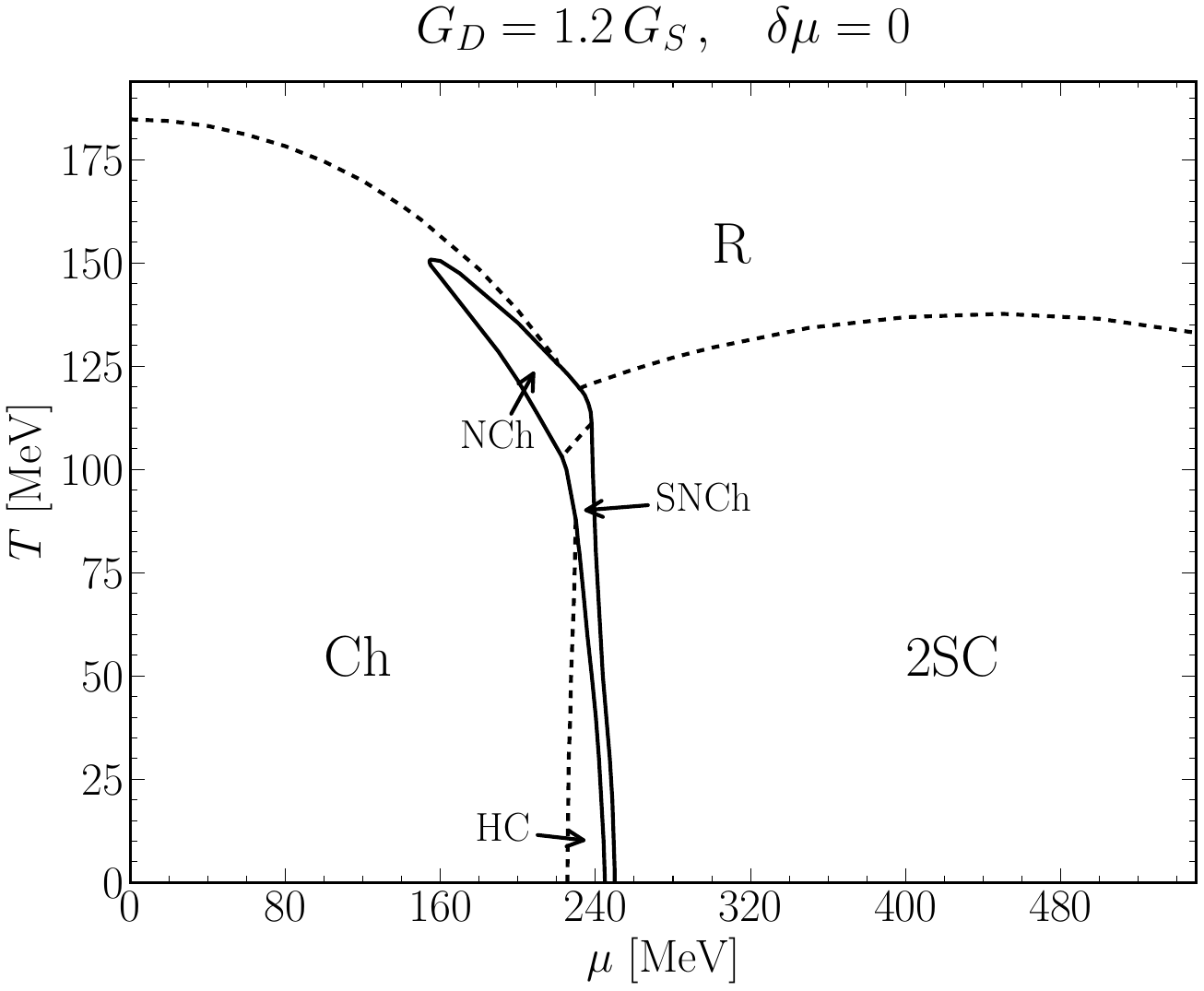}
\hfil
\includegraphics[width=8cm]{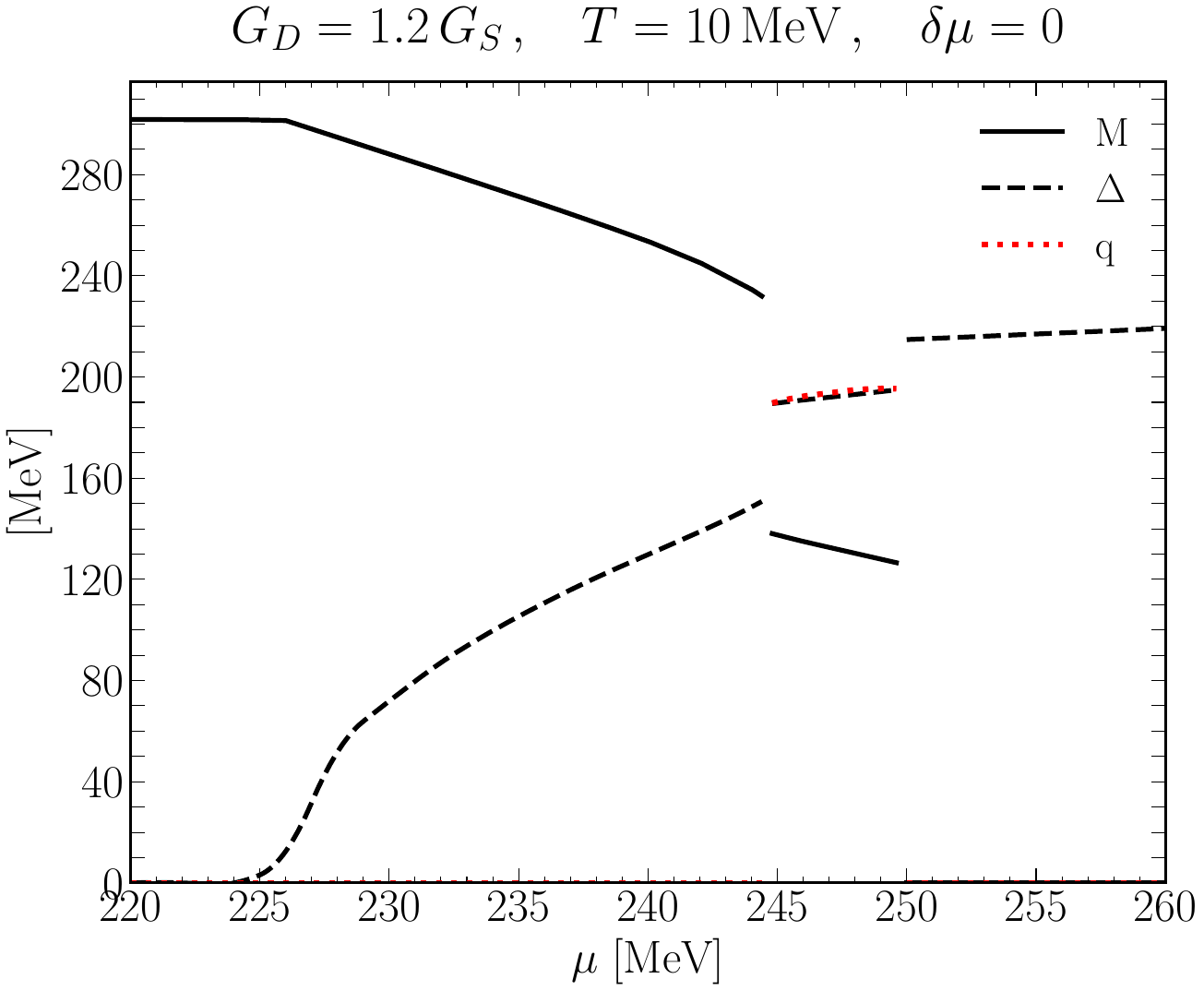}
\caption{Left: $T-\mu$ phase diagram for $G_D=1.2\,G_S$ at $\delta \mu=0$. Solid lines denote first-order, dashed lines second-order phase transitions.
Right: Corresponding order parameters at $T=10$~MeV as functions of the chemical potential $\mu$. 
\label{fig4-1}}
\end{center}
\end{figure*}

\section{Results}

In this section we present our numerical results. 
We adopt the parameter values of Ref.~\cite{Sadzikowski:2006jq}, $G_S=5.01\,\text{GeV}^{-2}$ for the quark-antiquark coupling constant and $\Lambda=0.65\,\text{GeV}$ for the momentum cutoff. In vacuum, this yields $F_{\pi}=0.093\,\text{GeV}$ for the pion decay constant, $M_0=0.301\,\text{GeV}$ for the constituent quark mass, and $\left\langle\bar{u} u\right\rangle=\left\langle\bar{d} d\right\rangle=(-0.25\,\text{GeV})^{3}$ for the chiral condensate per flavor.

The value of the quark-quark coupling constant $G_{D}$ is less constrained. Some authors relate it to the coupling constant $G_S$ via a Fierz transformation of a color current interaction, leading to $G_D = 0.75\,G_S$ \cite{BUBALLA2005205}, while fits to astrophysical observables yield values as large as $G_D \sim 1.5\,G_S$ or even higher~\cite{Christian:2025dhe}.
Here, in order to get a systematic understanding of the role of this coupling, we treat it as an independent parameter and vary $G_D/G_S$ between $0.3$ and $1.2$.

In the following, we investigate the phase structure and the corresponding order parameters as functions of the teperature $T$, the average chemical potential $\mu$ and the chemical potential difference $\delta\mu$. In our setup we have four different order parameters: the amplitudes $M$  and $\Delta$ of the chiral and the diquark condensate, and the corresponding wave numbers $q\equiv |\vec{q}|$ and $q'\equiv |\vec{q'}|$, respectively. Depending on which of these order parameters are non-vanishing, we can define nine distinct phases, which are summarized in Table \ref{table01}. (Note that non-vanishing wave numbers only make a physical difference if the corresponding amplitude is non-vanishing as well.) 
For most phases, we adopted the nomenclature of Ref.~\cite{Sadzikowski:2006jq}.

\subsection{The $T-\mu$  phase diagram
\label{Tmuphase}}

We begin with the discussion of the $T-\mu$  phase diagram, both for $\delta\mu = 0$ and $\delta\mu\neq 0$. We checked that for small and moderate values of the quark-quark coupling constant ($G_D/G_S \lesssim 1$) and $\delta\mu = 0$ our results are consistent with those of Ref.~\cite{Sadzikowski:2006jq}.
In particular we confirm the existence of an SNCh phase at intermediate chemical potential and not too high $T$, i.e., a phase where a CDW coexists with a homogeneous 2SC condensate. 
This phase is placed between a homogeneous chiral phase (Ch) at lower and a homogeneous 2SC phase at higher chemical potential, and an inhomogeneous chiral phase without 2SC condensates (NCh) at higher $T$.   

When we increase the quark-quark coupling constant, eventually a new phase occurs: As shown in Fig.~\ref{fig4-1} 
for $G_D=1.2\,G_S$, there is a region between the Ch and the SNCh phase where we find a HC phase, i.e., a phase where a homogeneous chiral condensate ($M\neq 0$, $\vec{q}=0$) coexists with a homogeneous 2SC phase ($\Delta \neq 0$, $\vec{q'}=0$). Such a phase has been reported in many previous papers \cite{Sadzikowski:2001th,Blaschke:2002xr,Park:1999bz,Rapp:2000zd,Schwarz:1999dj,Berges:1998rc,Vanderheyden:2000bz,Huang:2001yw}. 
The underlying physics is that, for sufficiently strong quark-quark coupling, bound diquark states exist in the model, which Bose condense when $2\mu$ exceeds their mass.\footnote{This holds strictly at $T=0$.} The phase transition from the Ch to the HC phase is therefore second order. 
On the other hand, the transition from a homogeneous chiral condensate to a CDW is always first order. Consequently, the HC-SNCh transition as well as the Ch-SNCh and the Ch-NCh transitions are first-order.

For illustration we show in the right panel of Fig.~\ref{fig4-1}, the order parameters as functions of $\mu$ at a fixed temperature $T=10$~MeV. 
Here the first-order transitions are visible as discontinuities, while the second-oder Ch-HC transition is continuous.
It would be interesting to see whether the CH-SNCh transition becomes continuous as well if instead of a CDW a ``real kink crystal'' is considered, which allows for a continuous transition between homogeneous and inhomogeneous chiral condensates~\cite{Thies:2006ti,Nickel:2009ke}. 
Such an analysis could be done along the lines of Ref.~\cite{PhysRevD.103.034030} but is beyond the scope of our paper.

The transition between the SNCh and 2SC is also first order. 
All other phase transitions, which are either related to the melting of the chiral condensate (Ch-R) or of the diquark condensate (SNCh-NCh and 2SC-R), are second order.

Finally, we note that the ``nose-like'' shape of the NCh phase, which was also found in Ref.~\cite{Sadzikowski:2006jq}, is an artifact of the regularization. As shown in Ref.~\cite{Nickel:2009ke}, the Ch, Nch and R phases should meet in a Lifshitz point which coincides with the tricritical point of the chiral phase transition in a homogeneous analysis. In fact, the ``nose'' is absent in the analysis of Ref.~\cite{PhysRevD.103.034030}, where Pauli-Villars regularization was used. Unfortunately, it is difficult to generalize this scheme to include LOFF phases. We therefore stick to the momentum cutoff, being aware of this issue. However, it would be interesting to investigate whether the cutoff artifacts can be mitigated by employing the renormalization-group consistent method discussed in Refs.~\cite{Braun:2018svj,Gholami:2024diy,Gholami:2024vly}, or through the ``Medium Separation Scheme" of Refs.~\cite{Farias:2005cr,Duarte:2018kfd}.

\begin{figure}[!htb]
\begin{center}
\includegraphics[width=8cm]{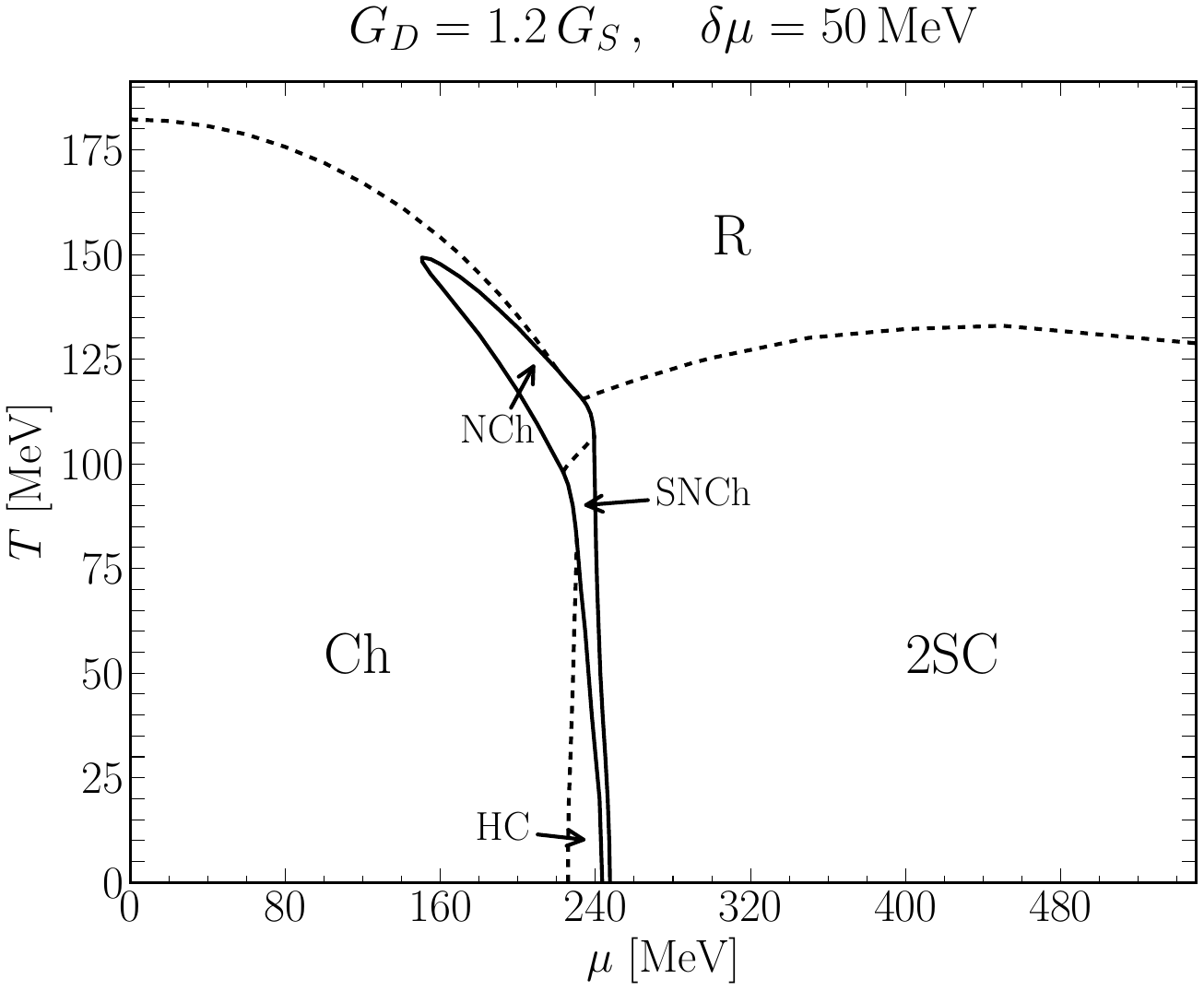}
\caption{The $T-\mu$ phase diagram for $G_D=1.2G_S$ at $\delta \mu=50$MeV.}
\label{fig4-3}
\end{center}
\end{figure}

In Fig.~\ref{fig4-3} we show the corresponding phase diagram for $\delta\mu = 50$~MeV. Comparing it with Fig.~\ref{fig4-1}, we see that the HC phase gets smaller but there are no qualitative and only small quantitative changes. In particular, we do not find that the chemical potential difference gives rise to the emergence of a LOFF phase. 
In fact, as we can see in the right panel of Fig.~\ref{fig4-1}, the gap at $T=10$~MeV, and hence at $T=0$, is larger than $\Delta = 200$~MeV. According to the Chandrasekhar-Clogston criterion we therefore expect the homogeneous 2SC phase to be stable until $\delta\mu \approx 150$ MeV for this coupling. 

\begin{figure*}[t]
 \centering
\includegraphics[width=7cm]{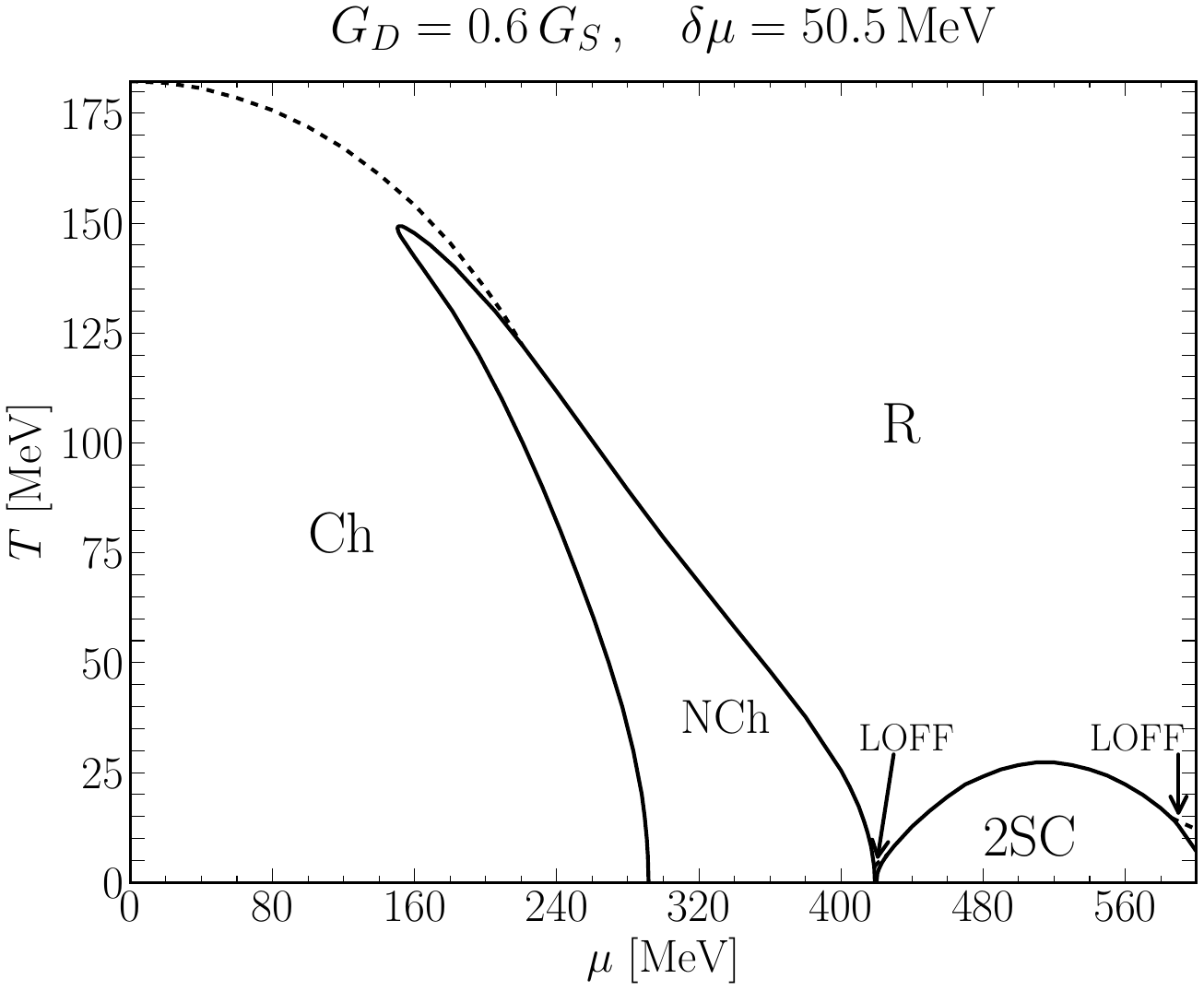}
\hfil
\includegraphics[width=7cm]{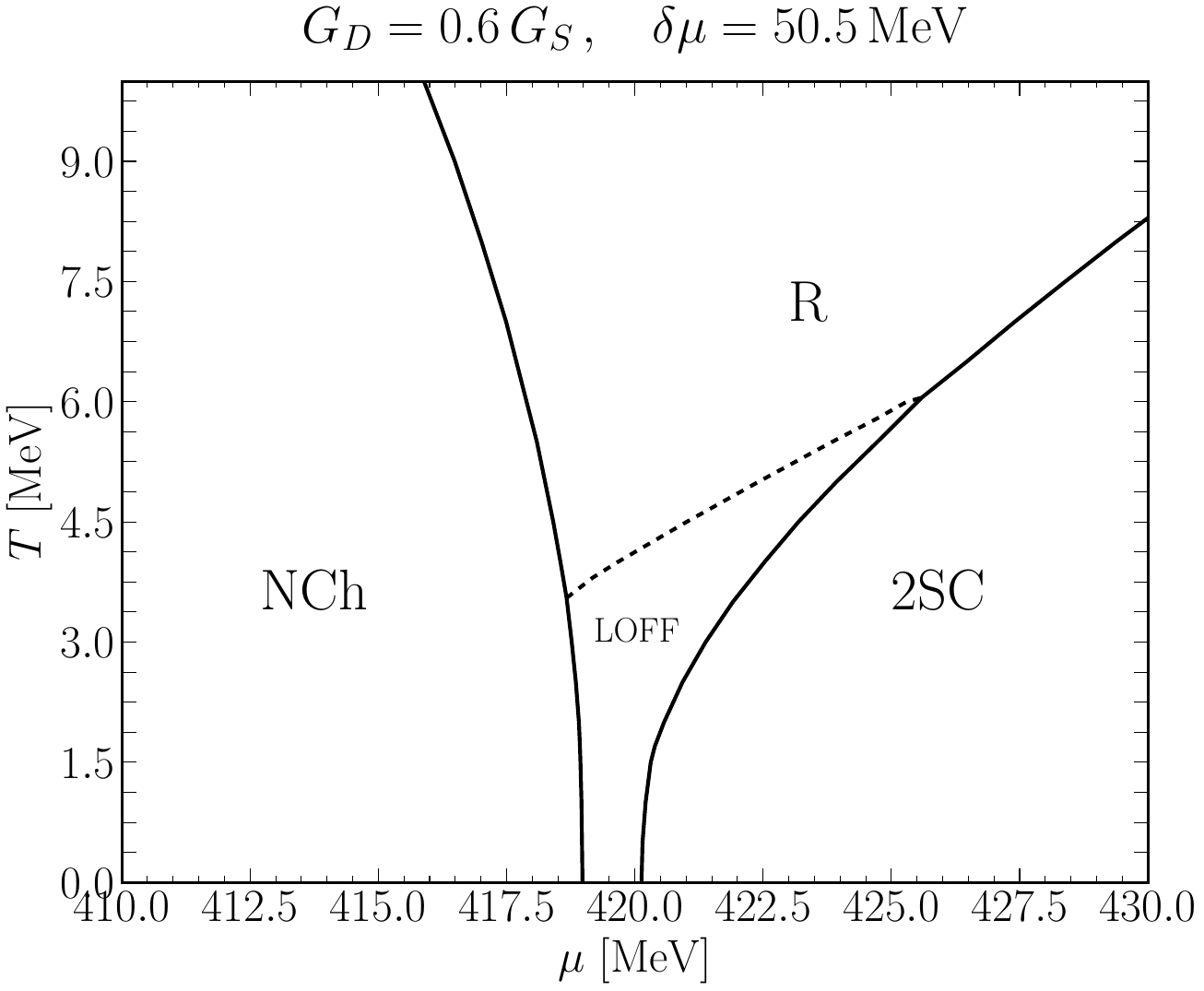}
\caption{The $T-\mu$ phase diagram for $G_D=0.6\,G_S$ at $\delta \mu=50.5$~MeV.
The right plot is a zoomed-in detail of the left one.
}
\label{phase5-3-2}
\end{figure*}

In this context we should note that in our analysis we did not consider charged pion condensates ($\pi_1 \neq 0$ or $\pi_2\neq 0$), which become favored over the scalar chiral condensate for $\delta\mu > m_\pi/2$ at low $T$~\cite{Barducci:2004tt,Warringa:2005jh}.
Strictly speaking, since we work in the chiral limit, we have $m_\pi = 0$, and therefore charged pion condensation should occur for any $\delta\mu \neq 0$.
Here we take the more phenomenological point of view that the neglect of charged pion condensates is justified as long as $\delta\mu$ is less than half of the physical pion mass. We therefore limit our analysis to $\delta\mu < 70$~MeV.
Then, in order to increase the chance of finding a LOFF phase in this range, we should decrease the quark-quark coupling, lowering the gap at $\delta\mu = 0$ and thus the Chandrasekhar-Clogston limit.

\begin{figure}[t]
  \centering
 \includegraphics[width=7cm]{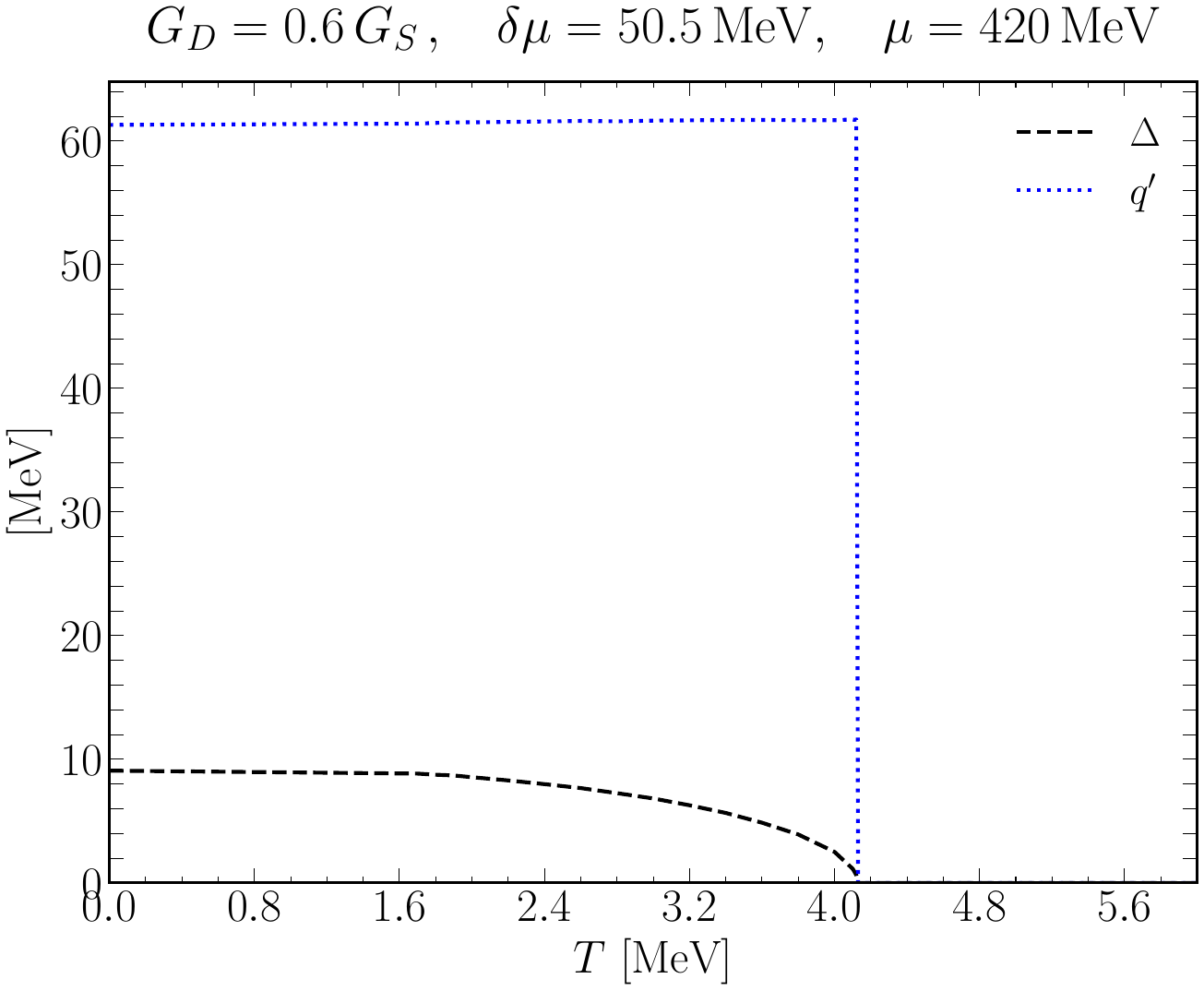}
 \caption{The order parameters as functions of $T$ at $G_D=0.6\,G_S$ and $\mu=420$~MeV and $\delta\mu=50.5$~MeV.
 \label{order5-3}}
 \end{figure}

Indeed, for $G_D = 0.6\,G_S$ we find a LOFF phase at values of $\delta\mu$ below half of the physical pion mass. This is shown in Fig.~\ref{phase5-3-2},
where the $T-\mu$ phase diagram is displayed for $\delta\mu = 50.5$~MeV. 
The right plot is a zoomed-in detail of the left plot, focusing on the region around $\mu=420$~MeV and low temperatures. 
The corresponding order parameters $\Delta$ and $q'$ are displayed as functions of $T$ at $\mu = 420$~MeV and $\delta\mu = 50.5$~MeV in  Fig.~\ref{order5-3}. We find $q' \approx 60$~MeV, corresponding to a LOFF phase with a wavelength of about $10$~fm. However, the amplitude is quite small and therefore the condensate melts already at a temperature slightly above 4~MeV. 
In the next subsection, we will therefore restrict ourselves to $T=0$ and analyze the phase structure in the $\mu-\delta\mu$ plane.

With increasing chemical potential, the 2SC condensate gets strengthened and therefore its critical temperature rises. As a consequence the LOFF phase ceases to exist above about $\mu = 426$~MeV, where the 2SC critical temperature overshoots the LOFF one, see right plot in Fig.~\ref{phase5-3-2}.

However, as one can see in the left plot of \cref{phase5-3-2}, eventually the 2SC critical temperature reaches a maximum and then decreases with increasing $\mu$, leading to the appearance of a second LOFF phase around $\mu = 600$~MeV. 
This is clearly a cutoff artifact: 
The decrease of the 2SC critical temperature can be traced back to the limited integration domain, which strongly suppresses the pairing when the chemical potential approaches the cutoff. Without this effect, the 2SC critical temperature would keep rising, leaving no room for a second LOFF regime. 
In principle the cutoff artifacts can be removed by applying one of the renormalization-group consistent regularization schemes introduced in Ref.~\cite{Gholami:2024diy}. Here we only note that our results should not be trusted in the region where the 2SC critical temperature decreases with increasing chemical potential, i.e., 
in the present case, above $\mu \approx 500$~MeV.

\begin{figure*}[t]
 \centering
\includegraphics[width=7cm]{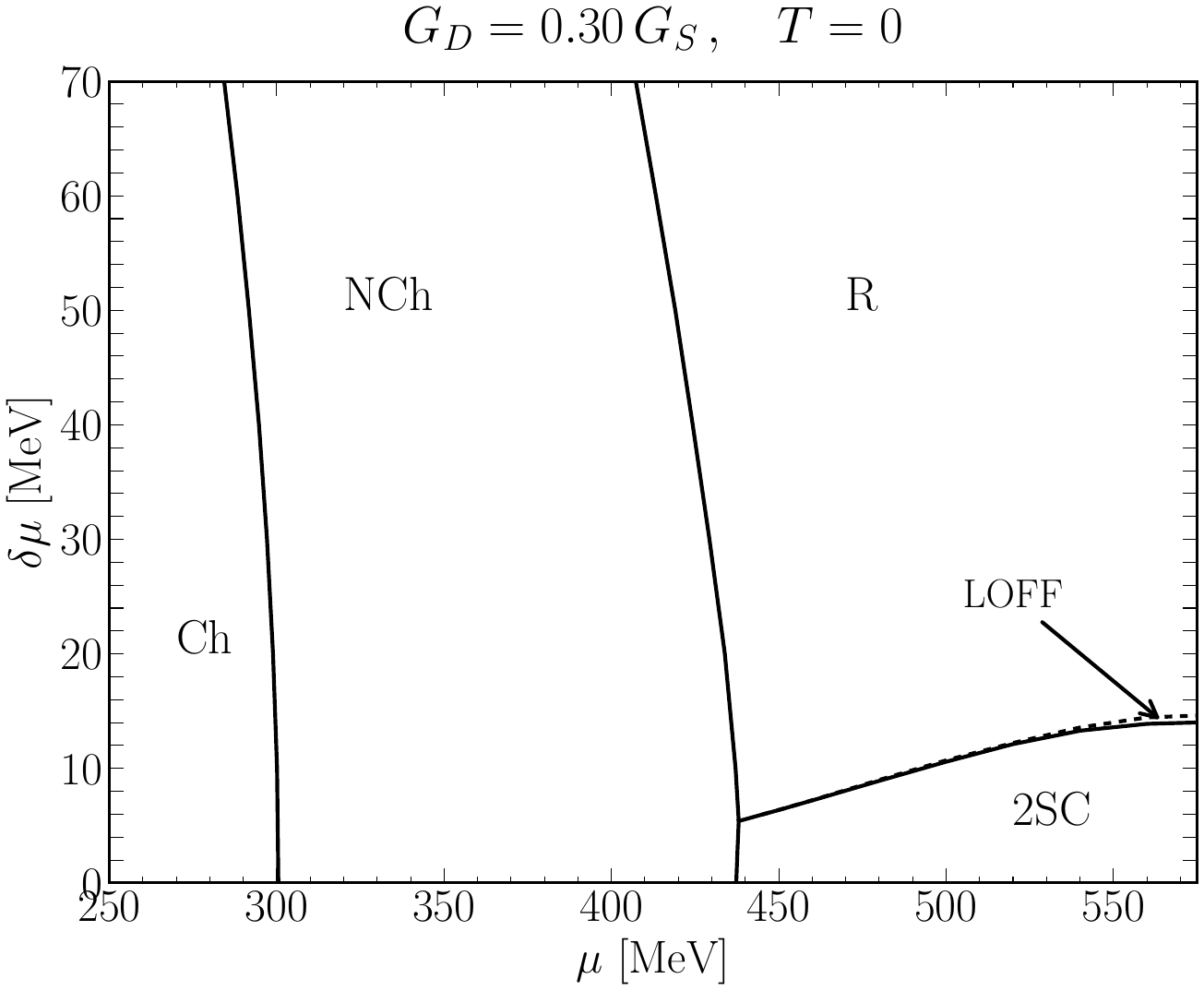}
\hfil
\includegraphics[width=7cm]{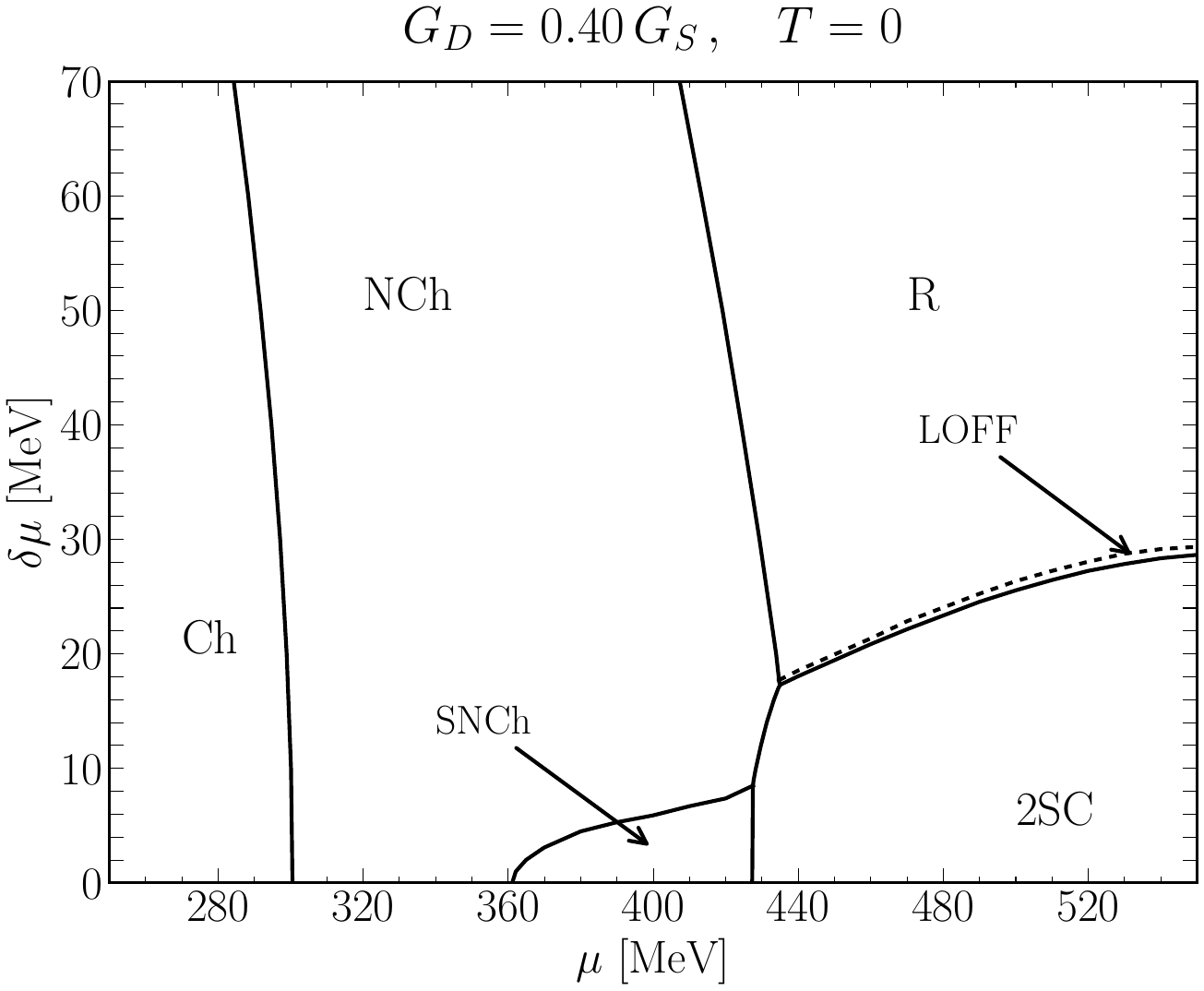}
\caption{$\mu-\delta\mu$ phase diagrams at $T=0$ for $G_D=0.3\,G_S$ (left) and $G_D=0.4\,G_S$ (right). Solid lines denote first-order, dashed lines second-order phase transitions.
\label{fig5-1}}
\end{figure*}

\begin{figure*}[t]
 \centering
\includegraphics[width=7cm]{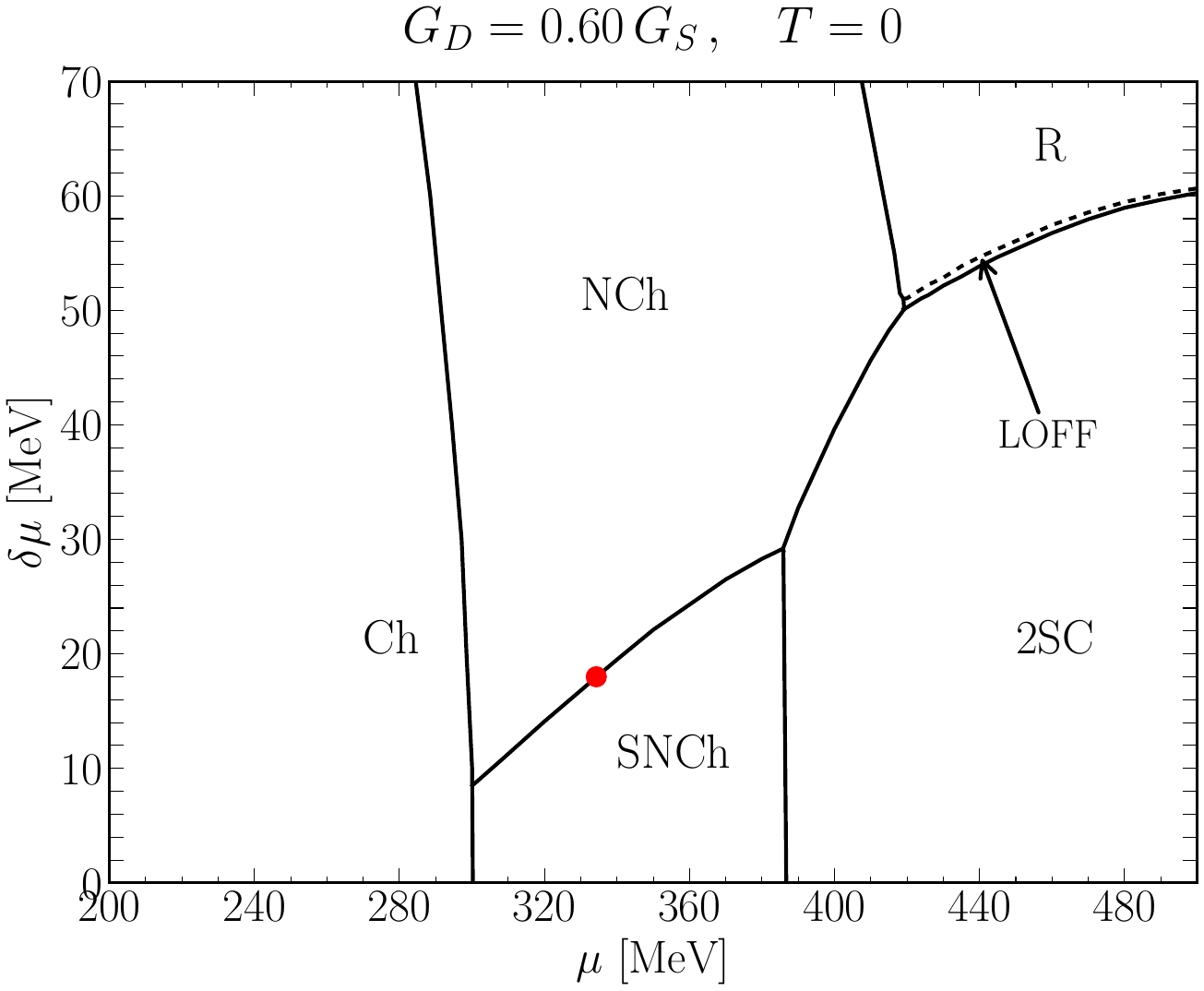}
\hfil
\includegraphics[width=6.6cm]{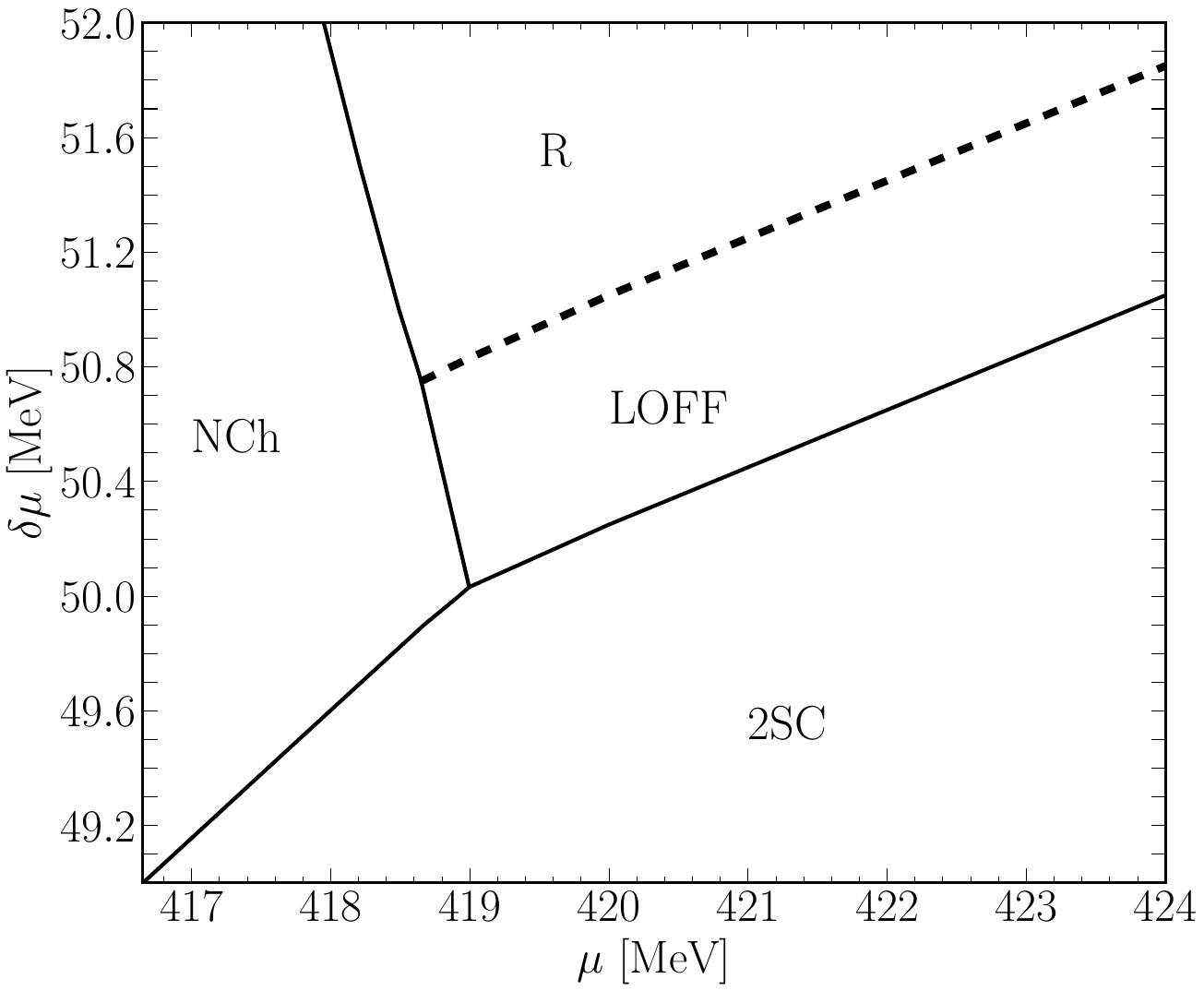}
\caption{$\mu-\delta\mu$ phase diagrams at $T=0$ for $G_D=0.6\,G_S$. 
Solid lines denote first-order, dashed lines second-order phase transitions.
The red dot indicates the point at which the stability analysis of \cref{omegaasmu} is performed.
The plot on the right-hand side is a zoomed-in detail of the left plot. 
\label{fig5-2}}
\end{figure*}

\subsection{The $\mu-\delta\mu$  phase diagram
\label{mudeltamuphase}}

In this subsection we investigate the $\mu-\delta\mu$ phase diagrams at $T=0$  
for different values of the quark-quark coupling constant $G_D$ in order to study its effect on the phase structure. 
As in the $T-\mu$ phase diagrams discussed in the previous subsection, the numerical results become severely affected by cutoff artifacts at large $\mu$. We therefore show the results only up to the value of $\mu$ where the critical $\delta\mu$ of the 2SC phase reaches a maximum. Note that this value depends on the diquark coupling and decreases with increasing $G_D$. 

In Fig.~\ref{fig5-1}, we show the phase diagrams for $G_D=0.3\,G_S$ (left) and $G_D=0.4\,G_S$ (right). 
In both cases we find a Ch phase at low $\mu$ and an NCh phase at intermediate $\mu$. At higher values of $\mu$ and small $\delta\mu$ there is a 2SC phase, which eventually gets unstable when $\delta\mu$ is increased. This gives rise to a LOFF phase in a small strip, followed by a non-superconducting chiral restored phase (R) at higher $\delta\mu$. 
For example, for $G_D=0.4\,G_S$ at $\mu=450$~MeV, we find the LOFF phase to be favored in the window $19.45$~MeV $ < \mu < 19.95$~MeV, corresponding to $0.684< \delta\mu/\Delta_0<0.701$, where $\Delta_0 = 28.44$~MeV is the 2SC superconducting gap parameter at $\delta\mu=0$. 
This window is lower and also more narrow than the weak-coupling prediction $0.707< \delta\mu/\Delta_0<0.754$ of Ref.~\cite{Bowers:2001ip}.\footnote{Using more sophisticated LOFF shapes than the single-plane-wave ansatz could increase this window by about a factor of 2~\cite{Nickel:2009ke}.}

In contrast, the NCh phase is quite insensitive against increasing the chemical potential difference and is stable for all values of $\delta\mu$ shown in the figure. 
All phase boundaries are of first order, except for the boundary between LOFF and R phase, which is second order.

Next, we focus on the effect of an increasing quark-quark coupling on the phase structure.
Comparing the left phase diagram of Fig.~\ref{fig5-1} with the right one, we see that, with increasing $G_D$, an SNCh phase emerges, which is not present at smaller couplings.  
At $G_D = 0.4\,G_S$ (right plot Fig.~\ref{fig5-1}), the SNCh phase is connected to the 2SC phase at the upper-$\mu$ end but does not reach down to the Ch phase at lower $\mu$, so that the NCh phase is still present at $\delta\mu = 0$ in a certain chemical-potential interval.
However, when $G_D$ is increased further, the SNCh phase expands and  eventually connects to the Ch phase, as shown in Fig.~\ref{fig5-2} for $G_D = 0.6\,G_S$.

\begin{figure*}[t]
 \centering
\includegraphics[width=7cm]{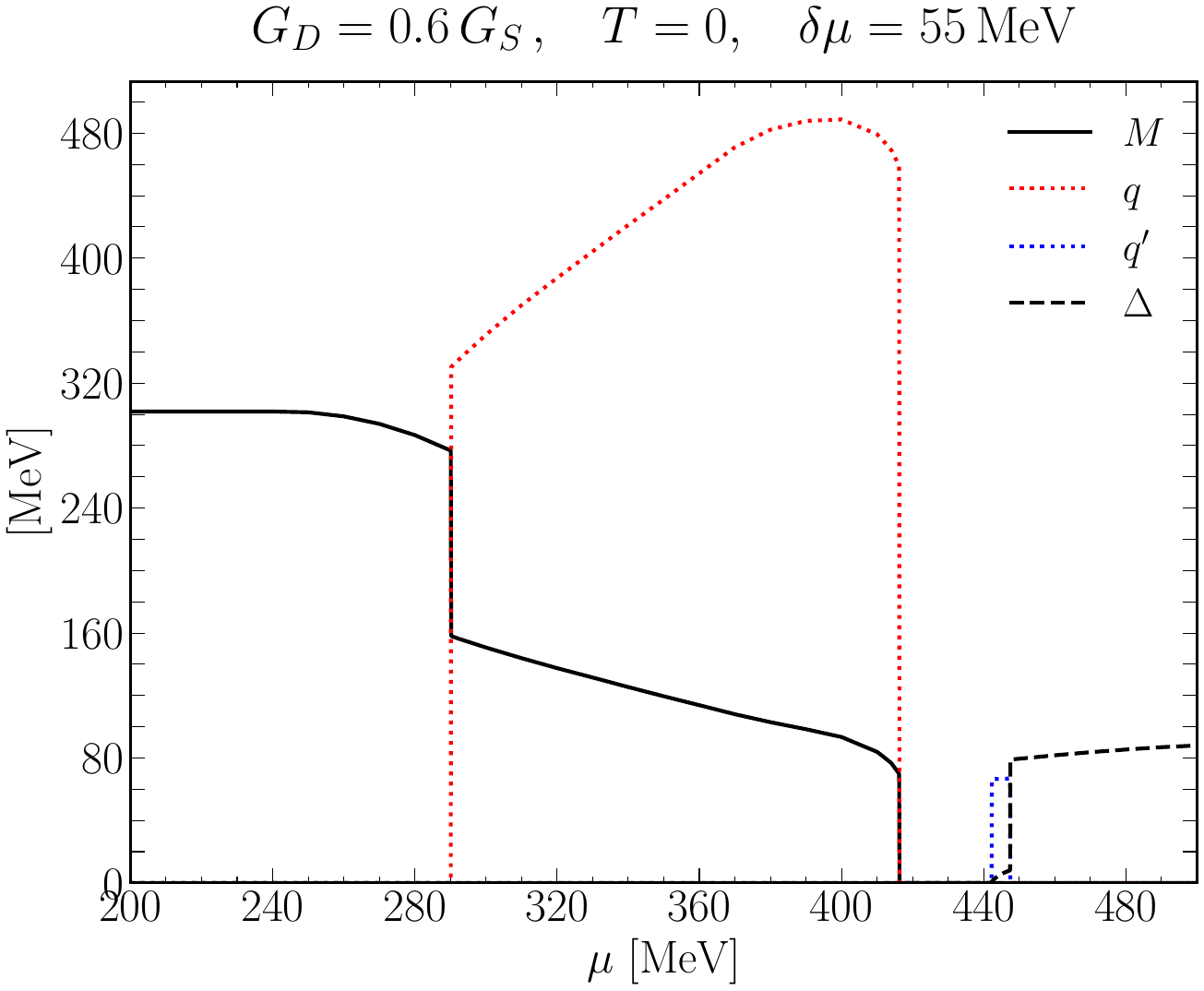}
\hfil
\includegraphics[width=7cm]{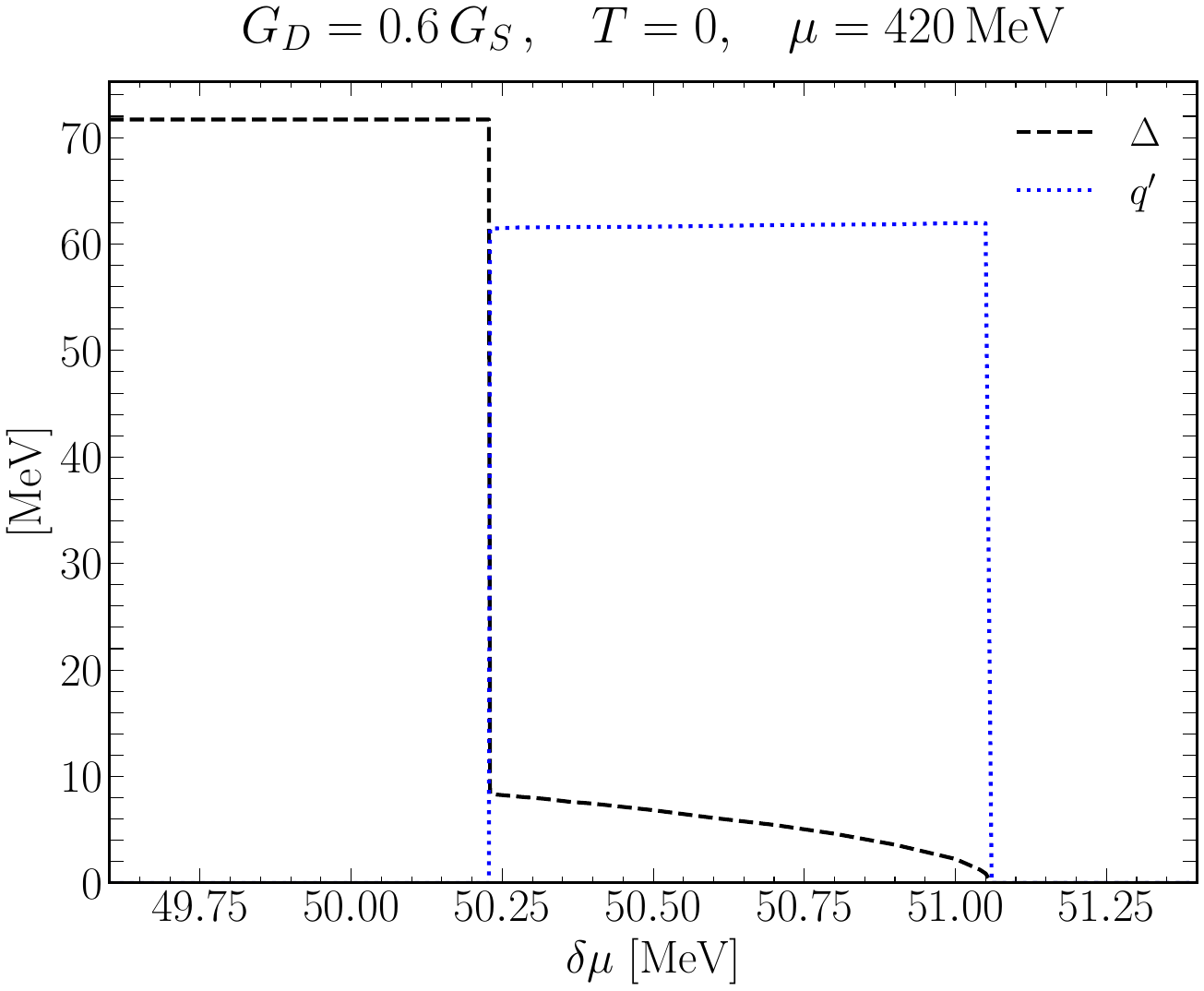}
\caption{Order parameters at $T=0$ for $G_D=0.6\,G_S$ 
as functions of $\mu$ at $\delta\mu=55$MeV (left) and as functions of $\delta\mu$ at $\mu=420$MeV (right).
\label{order5-1}}
\end{figure*}

Our main result is that we do not find any region of the phase diagram where $\vec q$ and $\vec q\,'$ are simultaneously nonzero, i.e., where a CDW coexists with a LOFF phase. This is exemplified by the $\mu-\delta \mu$ phase diagram for $G_D=0.6\,G_S$ in Fig.~\ref{fig5-2}.
Further insights about the nature of the phase transitions can be obtained from Fig.~\ref{order5-1}, where the order parameters are displayed as functions of $\mu$ along a line of constant $\delta \mu$ and vice versa.

As one can see in the right plot of Fig.~\ref{fig5-2}, the NCh phase, where $\vec q \neq 0$, shares a first-order phase boundary with the LOFF phase, where $\vec q\,' \neq 0$, but there is no region where the two inhomogeneities coexist. In fact, since the amplitude $\Delta$ of the diquark gap vanishes in the NCh phase, while the mass amplitude $M$ vanishes in the LOFF phase, the transition between them corresponds to a rather drastic change of the phase structure.

\begin{figure}[hbt]
\centering
\includegraphics[width=7cm]{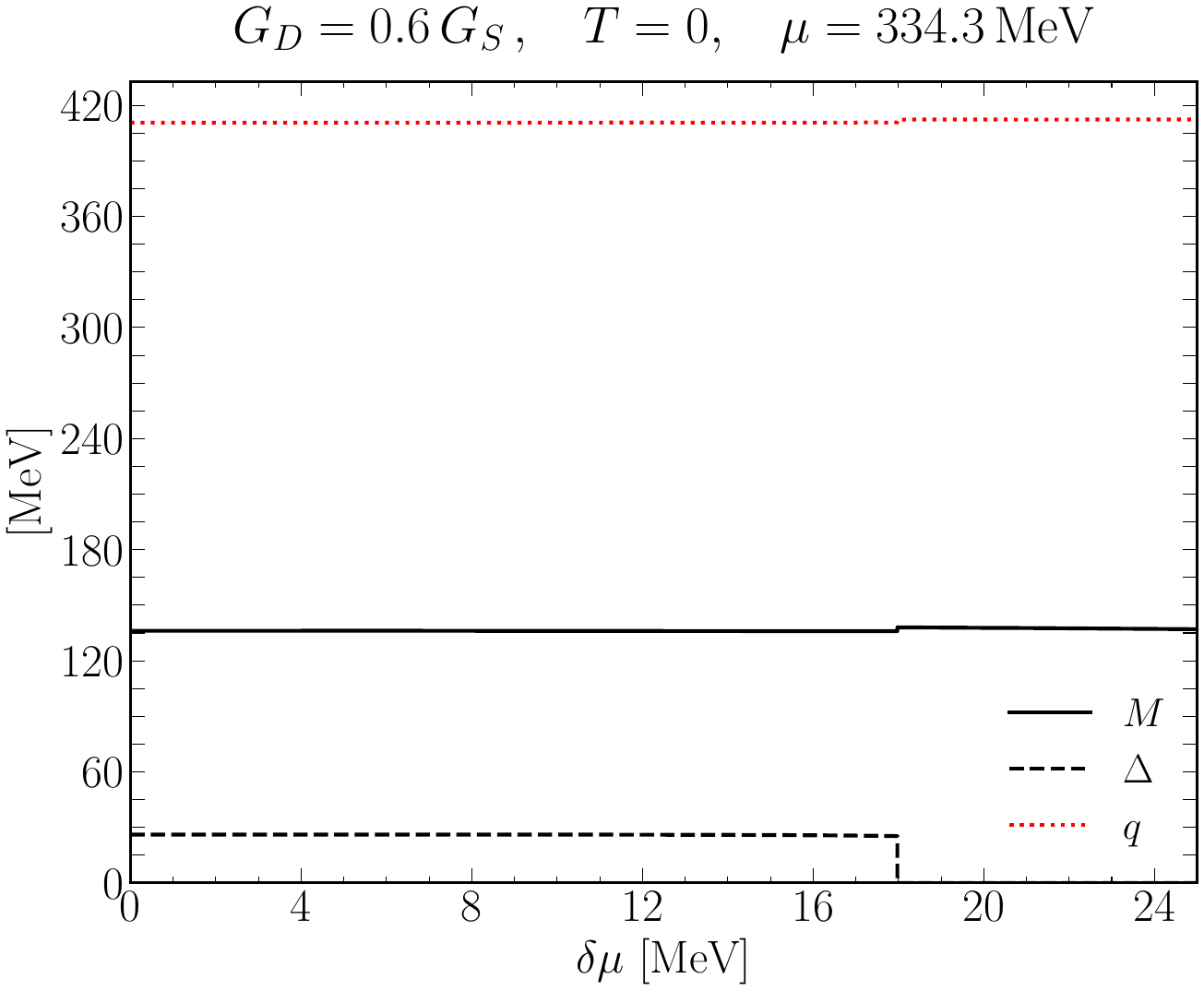}
\caption{The order parameters as function of $\delta\mu$ at $G_D=0.6\,G_S$ and $\mu=334.3$ MeV. The transitions between SNCh and NCh phases is first order. Notably, the order parameters $M$ and $q$ exhibit only minor changes across the phase transition.}
\label{orderasmu1}
\end{figure}

\begin{figure}[!htb]
\centering
\includegraphics[width=7cm]{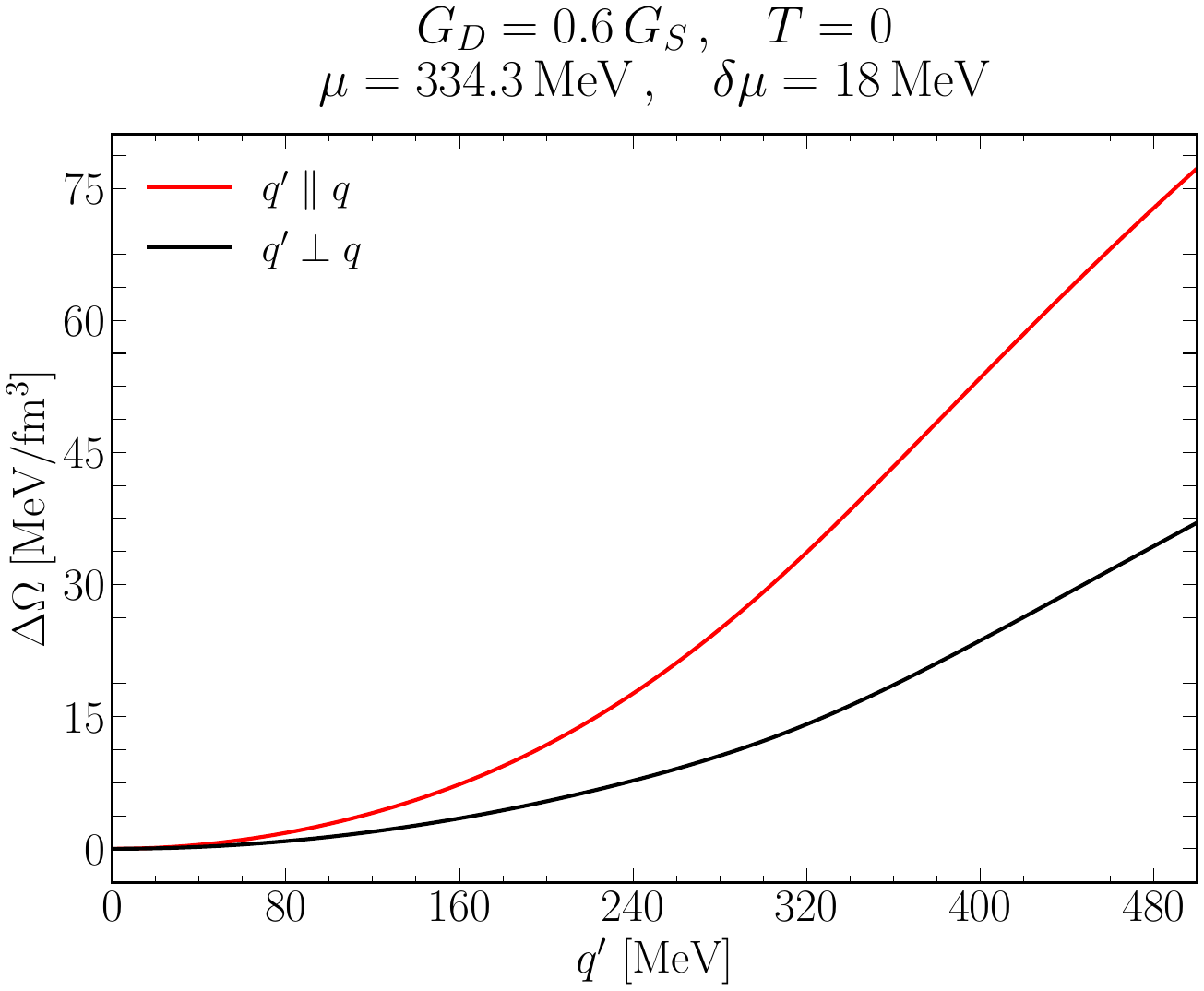}
\caption{
Excess effective potential per unit volume, $\Delta\Omega(q') = \Omega(q')-\Omega(0)$,
as a function of $q^{\prime}$
for a CDW with 
fixed $M=138$~MeV and $q=412$~MeV together with a LOFF-like modulation with $q'$
at $G_D=0.6\,G_S$, $T=0$, $\mu=334.3$~MeV and $\delta\mu=18$~MeV (red dot in \cref{fig5-2}).
For each value of $q'$,
the value of $\Delta$ is determined by minimization. 
Parallel and antiparallel configurations lie on top of each other.}
\label{omegaasmu}
\end{figure}

        Therefore, it is probably more likely to find a phase where CDW and LOFF modulations coexist in a region adjacent to the SNCh phase, where already a homogeneous 2SC condensate exists in coexistence with a CDW. In particular it seems conceivable that such a CDW-LOFF coexistence phase exists in a thin strip between SNCh and NCh phase, similar to the LOFF window between 2SC and restored phase. 
        We investigated this possibility carefully, including different relative orientations between the two wave vectors $\vec q$ and $\vec q\,'$ However, we never found a configuration that is energetically favored. 
        In order to illustrate this and to gain further insights, we performed a pseudo-stability analysis for a point directly at the SNCh--NCh phase boundary. To be specific, we      
        selected the point $\mu = 334.3~\mathrm{MeV}$ and $\delta\mu = 18~\mathrm{MeV}$, which is indicated by a red dot in \cref{fig5-2}. 
The strategy is the following: As can be seen from
\cref{orderasmu1}, $M$ and $q$ do not change strongly across the phase boundary. Hence we fix their values to those on the NCh-side of the boundary.
Then we vary $q'$ and, for each value of $q'$, we minimize the effective potential with respect to $\Delta$.
If a LOFF instability was present, the potential should develop a negative curvature and favor a finite value of $q'$. In that case $\Delta\Omega(q')\equiv\Omega(q')-\Omega(0)$ decreases from zero with increasing $|q'|$, reaching a minimum at a nonzero $q'$. In \cref{omegaasmu}, we show the results of this analysis for two representative relative orientations between $\vec q$ and $\vec q\,'$: parallel and perpendicular.\footnote{It can be shown that the result is an even function of $\vec q\cdot\vec q\,'$, so parallel and antiparallel orientations yield identical results.} 
In all cases, the potential remains stable around $\vec q\,' = 0$, indicating that no LOFF window above the SNCh phase is favored in \cref{fig5-2}.

We also note that the SNCh phase and the LOFF phase never have a common phase boundary. 
Their closest points are two triple points, corresponding to the upper and lower end of a first-order 2SC-NCh phase boundary. 
It turns out that it is not possible to have the two triple points coincide or even get a common SNCh-LOFF phase boundary by varying the value of $G_D$. This can be seen by comparing Fig.~\ref{fig5-1} with Fig.~\ref{fig5-2}: While increasing the value of $G_D$ from 
$0.4\,G_S$ to $0.6\,G_S$ shifts the triple point between SNCh, 2SC and NCh phase to higher values of $\delta\mu$, the triple point between 2SC, NCh and LOFF phase moves to higher $\delta\mu$ as well, and their distance even gets larger.

\section {Summary and discussion}
\label{s5}

We have studied the phase diagram of two-flavor quark matter within the Nambu--Jona-Lasinio (NJL) model in the chiral limit, incorporating simultaneously the chiral density wave (CDW) and the single-plane-wave Ovchinnikov-Fulde-Ferrell (LOFF) phase of color superconductivity, treated as independent variational degrees of freedom. Using the three-momentum cutoff regularization scheme~\cite{Sadzikowski:2002iy}
with the subtraction procedure of Ref.~\cite{He:2006vr}, we minimized the thermodynamic potential with respect to all order parameters, i.e., the chiral amplitude $M$, the diquark amplitude $\Delta$, and both modulation wave vectors $\vec{q}$ and $\vec{q'}$, without imposing any constraint on their relative orientation.

Our central result is the absence of any phase in which both 
$\vec{q} \neq 0$ and $\vec{q'} \neq 0$ simultaneously. While inhomogeneous chiral and diquark condensates meet at a first-order phase boundary between NCh and LOFF phase, they are mutually exclusive 
across the entire parameter space explored, independently of 
the relative orientation of $\vec{q}$ and $\vec{q'}$.
This was 
further confirmed by a stability analysis at the SNCh-NCh phase 
boundary, where the effective potential shows no instability 
toward finite $|\vec{q'}|$. We note that the
inhomogeneous chiral 
condensate, however, can coexist with the 2SC phase, as reported earlier in Refs.~\cite{Sadzikowski:2002iy,Sadzikowski:2006jq,PhysRevD.103.034030}.

In the $T$-$\mu$ phase diagram at $\delta\mu = 0$, our results are 
consistent with those of Ref.~\cite{Sadzikowski:2006jq} for moderate 
diquark couplings. For sufficiently large $G_D$, we additionally 
identify a homogeneous coexistence phase in which a homogeneous 
chiral condensate coexists with the 2SC phase. For smaller couplings, 
where the Chandrasekhar-Clogston limit lies within the accessible 
range of $\delta\mu$, a narrow LOFF window appears at low 
temperatures. In the $\mu$-$\delta\mu$ plane at $T = 0$, increasing 
$G_D$ drives the emergence of a phase where a CDW coexists with a 
homogeneous 2SC condensate
(SNCh), while the LOFF phase persists as a 
narrow strip at the boundary of the 2SC phase at high $\delta\mu$.

Several directions remain open for future work. The cutoff artifacts 
identified here call for renormalization-group consistent 
regularization~\cite{Gholami:2024diy,Braun:2018svj}. It would also be 
interesting to investigate whether more general spatial modulations 
beyond the single plane-wave ansatz~\cite{Nickel:2009ke,Thies:2006ti} 
could open a coexistence window. Another natural extension of the present work would be a dedicated stability
analysis of the stationary solutions found here. In particular, it would
be interesting to apply the methodology developed in
Ref.~\cite{Motta:2023pks} to investigate the stability properties of the
various homogeneous and inhomogeneous phases. Finally, extending this analysis to 
include pion condensation, and applying the charge neutrality and $\beta$-equilibrium would be an 
important step toward applications to neutron star matter.
\medskip
\medskip

\section*{Data Availability}

The numerical data presented in all figures in this work are openly
available in the ancillary files of the corresponding arXiv
submission.

\acknowledgments{
C.F. Mu is grateful to Dr. Xuewen Hao and Prof. Lianyi He for helpful
discussions. C.F. Mu also acknowledges the financial support from the China Scholarship Council (CSC No. 202008330252). This work is supported by the National Natural Science Foundation of China under Grant No. 12275081 and the Natural Science Foundation of Zhejiang Province under grant
No. LY19A050002.  H. Gholami and M. Buballa thank the Deutsche Forschungsgemeinschaft
(DFG) through the Collaborative Research Center TransRegio CRC-TR 211 “Strong-interaction matter under extreme
conditions” and the individual grant FI 970/16-1.
\\
}

\onecolumngrid
\appendix

\bibliography{bib}

\end{document}